\documentclass[10pt]{IEEEtran}
\usepackage{graphicx}
\graphicspath{{figures/}}
\usepackage{amssymb}
\usepackage{amsmath}
\usepackage{epstopdf}
\usepackage[utf8]{inputenc}
\usepackage[T1]{fontenc}
\DeclareUnicodeCharacter{2194}{\ensuremath{\leftrightarrow}}

\usepackage{courier}
\usepackage{color}
\pagecolor{white}

\usepackage{fvextra}

\usepackage{parskip}
\usepackage{xurl}
\usepackage{enumitem}
\usepackage[table]{xcolor}

\usepackage{subcaption}
\usepackage{hyperref}
\usepackage{tablefootnote}
\usepackage{threeparttable}
\usepackage[numbers]{natbib}

\usepackage{listings}

\lstdefinelanguage{logsources}{
keywords=[1]{CARRIER_SERVICES, ANDROID_DIALER,SCOOBY_EVENT,ANDROID_MESSAGING}
}

\lstdefinelanguage{pseudo}{
    keywords=[1]{androidID,pseudonymousIdToken,NIDCookie,logSourceName,incomingPhoneNumber},
    keywords=[2]{header,deviceStatus,deviceInfo,dynamiteModules,accessibilitySettings,googlePlayServicesVersionName,sdkExtensions,logEntry,event},
    keywords=[3]{SDKVersion,currentTimeMillis,qosTier,elapsedTime,model,product,buildID,hardware,device,simOperator,language,country,manufacturer,brand,board,radioVersion,buildFingerprint,buildType,simCarrierId,timestamp,networkConnected,batteryCharged,accountIsGoogleCom,googlePlayServicesVersionString,googlePlayServicesVersionCode,packageVersionName,packageName},
}
\lstdefinelanguage{pseudo2}{
    keywords=[1]{AOSPEventType,timestamp,callDuration,callID},
    keywords=[2]{logEntry,event,experimentTokens,tokens,connectionDetail,deviceDetails,impressionEvent,connectionDetails,searchQuery,callDetails,callSetupTiming},
    keywords=[3]{buildDevice,buildModel,buildID,systemApp,deviceClass,token,tokenBytes,hasContent,bootCount,elapsedRealtime,subEvent,tz_offset,buildVersionRelease,releaseStatus,activeNetworkType,updatedSystemApp,simOperator,mobileOperator,country,installedBy,callInitiationType,isIncoming,lookupResultType,disconnectCause,queryLength,searchTimeMillis,elapsedDaysSinceDialerInstall},
}
\lstdefinelanguage{pseudo3}{
    keywords=[1]{eventType,timestamp,sha256HashMsg,sha256HashPrevMsg,currentTime_ms,elapsedTimeSinceMsgSendRecv_ms},
    keywords=[2]{logEntry,event,experimentTokens,tokens,bugleMessage,messageProtocol,bugleMessageStatus,subscriptionUtils,mimeData,messageTiming,phoneNumberFormat2,carrierServicesData,messageSettings,mimeData2,connectionDetails},
    keywords=[3]{subEvent,tz_offset,bootCount,elapsedRealtime,activeNetworkType,messageSendClickToSentLatency,bugleMessageSource,usageStatsLoggingId,conversationType,sendAttempt,wasRCS,rcsStatus,configStatus,elapsedTimeEnum,RcsConfig,messageStatus,conversationIdSHA1,Mcc,Mnc,simOperatorName,networkOperatorName,simCarrierId,mimeType,valid,messagePart,versionCode,hasParticpants,hasConversationId,sendModeIsAuto,sendModeIsXms,onlineStatus,sendmode,token,tokenBytes,hasContent}
}

\lstdefinelanguage{http}{
    alsoletter=-,
    keywords=[1]{DSID},
    keywordstyle=[2]\color{blue}\bfseries,
    keywords=[2]{},
    keywordstyle=[3]\color{brown},
    keywords=[3]{authorization, Authorization, user-agent, User-Agent, x-server-token, set-cookie, device:, x-goog-spatula, x-ps-rh, X-Android-Package},
    keywordstyle=[4]\color{blue}\bfseries,
    keywords=[4]{POST, GET},
}
\lstdefinelanguage{dialer}{
   basicstyle=\tiny\tt\color{brown},
    alsoletter=_,
    keywords=[1]{},
    keywords=[3]{callDuration},
    keywords=[2]{MAIN_CLICK_FAB_TO_OPEN_DIALPAD,queryLength,PRECALL_INITIATED,DIALER_GATEWAY_CALL_INFO_LOOKUP_SERVICE_PHONE_LOOKUP_UNAUTHENTICATED,DIALER_GATEWAY_CALL_INFO_LOOKUP_SERVICE_PHONE_LOOKUP_FAILED,TIDEPODS_STATUS_BAR_NOTIFICATION_SHOWED,TIDEPODS_BUBBLE_SHOWED,USER_PARTICIPATED_IN_A_CALL,ANNOTATED_CALL_LOG_FORCE_REFRESH_CHANGES_NEEDED,ANNOTATED_CALL_LOG_FORCE_REFRESH_NO_CHANGES_NEEDED, ANNOTATED_CALL_LOG_NOT_DIRTY,CALL_SCREENING_SERVICE_MUSIC_IS_NOT_ACTIVE,INCOMING_CALL_SCREENED,INCOMING_NON_SPAM_CALL,DIALER_GATEWAY_CALL_INFO_LOOKUP_SERVICE_PHONE_LOOKUP_FAILED,DARK_MODE_SET_FOLLOW_SYSTEM_BY_CONTACTS,IN_CALL_DIALPAD_HANG_UP_BUTTON_PRESSED,NON_SPAM_NOTIFICATION_NOT_SHOWN_AFTER_THROTTLE},
}
\lstdefinelanguage{messages}{
   basicstyle=\tiny\tt\color{brown},
    alsoletter=_,
    keywords=[3]{sendAttempt,conversationIdSHA1},
    keywords=[1]{sha256HashMsg,sha256HashPrevMsg,simSerialNumber},
    keywords=[2]{BUGLE_MESSAGE,BUGLE_APP_CONFIGURATION,BUGLE_COMPOSE,BUGLE_P2P_SUGGESTION,BUGLE_SPAM_DETECTION,BUGLE_TELEPHONY_EVENT,BUGLE_APP,BUGLE_CONVERSATION,BUGLE_CONTACT_BANNER},
}
\lstdefinelanguage{analytics}{
   basicstyle=\tiny\tt\color{brown},
    alsoletter=-,
    keywords=[1]{event_timestamp,google_ad_id,firebase_instance_id,package_name},
    keywords=[2]{bundle,event_info,user_info,event},
    keywords=[3]{message_timestamp,bundle_end_timestamp,last_bundle_end_timestamp,operating_system,operating_system_version,Build_MODEL,language_country,timezone_offset_mins,app_version,gmp_version,gms_version,random_hex,dev_cert_hash,daily_conversions_count,gmp_app_id,last_bundle_start_timestamp2,firebase_instance_id,app_version_int,config_version,M,dynamite_version},
}
\lstdefinelanguage{analytics_event}{
   basicstyle=\tiny\tt\color{brown},
    alsoletter=-,
    keywords=[1]{event_timestamp,event_code},
    keywords=[2]{event, event_info},
    keywords=[3]{setting_code,data_str,previous_event_timestamp},
}
\lstdefinelanguage{analytics_dialer}{
   basicstyle=\tiny\tt\color{brown},
    alsoletter=-,
    keywords=[3]{event_code,data_str},
}

\definecolor{synKw}{HTML}{1F4E9C}
\definecolor{synType}{HTML}{1B7F79}
\definecolor{synFn}{HTML}{B35900}
\definecolor{synNum}{HTML}{C0392B}
\definecolor{synIvar}{HTML}{7A3E9D}
\definecolor{synCmt}{HTML}{6B6B6B}
\newcommand{\synnum}[1]{{\color{synNum}\bfseries #1}}

\lstdefinelanguage{pbobjc}[Objective]{C}{
  morekeywords=[2]{NSString,PBCodable,PBDataWriter,int32_t,super},
  morekeywords=[3]{PBWriteStringField,PBWriteInt32Field,PBWriteFloatField},
  morekeywords=[4]{_first,_second,_third,_has,has_first,has_second,has_third},
  alsoletter={_@},
}

\lstdefinelanguage{armasm}{
  morekeywords=[1]{ldrsw,ldrb,ldr,tbnz,tbz,mov,bl,b},
  morekeywords=[2]{x0,x1,x8,x19,x20,x21,w1,w2,w8,s0},
  morekeywords=[3]{_PBWriteStringField,_PBWriteInt32Field,_PBWriteFloatField},
  morekeywords=[4]{L_first,L_second,L_third,L_epilogue},
  morekeywords=[5]{_has,_first,_second,_third},
  alsoletter={_},
  comment=[l]{;},
}

\lstdefinestyle{codefig}{
  escapechar={}, columns=fixed, keepspaces=true, breaklines=false,
  basicstyle=\scriptsize\ttfamily, linewidth=\linewidth,
  xleftmargin=2pt, xrightmargin=2pt, aboveskip=0pt, belowskip=0pt,
  frame=single, framerule=0.3pt, rulecolor=\color{gray!60},
  backgroundcolor=\color{gray!4},
  commentstyle=\color{synCmt}\itshape, identifierstyle=,
  keywordstyle=[1]\color{synKw}\bfseries,
  keywordstyle=[2]\color{synType}\bfseries,
  keywordstyle=[3]\color{synFn}\bfseries,
  keywordstyle=[4]\color{synIvar},
  keywordstyle=[5]\color{synIvar},
}
\lstdefinestyle{objcfig}{style=codefig, language=pbobjc,
  morecomment=[l]{//},
  literate={5)}{{\synnum{5})}}2 {2)}{{\synnum{2})}}2 {9)}{{\synnum{9})}}2,
}
\lstdefinestyle{asmfig}{style=codefig, language=armasm,
  literate={\#0x9}{{\synnum{\#0x9}}}4 {\#0x5}{{\synnum{\#0x5}}}4 {\#0x2}{{\synnum{\#0x2}}}4
           {\#ivar_off}{{\color{synKw}\#ivar\_off}}9,
}

\lstdefinelanguage{pbproto}{
  morekeywords=[1]{message},
  morekeywords=[2]{string,int32,float},
  morekeywords=[4]{first,second,third},
  alsoletter={_},
}
\lstdefinestyle{protofig}{style=codefig, language=pbproto,
  literate={5;}{{\synnum{5};}}2 {2;}{{\synnum{2};}}2 {9;}{{\synnum{9};}}2,
}

\usepackage{booktabs, tabularx, array, pdflscape, url}
   \usepackage{booktabs}
   \usepackage{array}
   \usepackage{multirow}
   \usepackage{longtable}  %
   \usepackage{textcomp}   %
   \usepackage{amssymb}    %
   \usepackage{xspace}

\providecommand{\tightlist}{%
  \setlength{\itemsep}{0pt}\setlength{\parskip}{0pt}}

\title{Decoding What iOS Location Services Send to Apple: An LLM Is an Unreliable Analyst but a Useful Toolmaker}
\author{D.J.Leith\thanks{Contact email: doug.leith@tcd.ie}\\Trinity College Dublin, Ireland}

\begin{document}
\maketitle

\begin{abstract}
Reverse engineering the data formats used by mobile software is skilled and slow, and using AI to assist with it is attractive. We evaluate this for the protobuf messages that an iPhone sends to Apple's location services. Working from Apple's binaries, we use two frontier models (Claude Opus 4.8, Codex GPT Sol 5.6) to reconstruct the schemas of protobufs observed in network traffic. The great majority of the schemas contain substantial errors, and these are hard to detect because the schemas are internally consistent. Direct AI use is therefore unsuitable for forensic and privacy analysis. AI can, however, write reliable analysis tools when their output can be validated. We use Claude to write a Python tool, Extractor, that reliably recovers protobuf schemas from Apple binaries. We validate it against binaries compiled from schemas with known answers, previously published schemas, and network measurements. We demonstrate the use of Extractor to decode and analyse the location services network traffic of a jailbroken iPhone 8 (iOS 16.7.11).

\end{abstract}
\section{Introduction}\label{introduction}

Mobile location data is sensitive. Knowing where a person is at night and during the day usually reveals where they live and work, and visits to a clinic, a place of worship or a protest can reveal health, belief and political activity. Handset collection of location data has repeatedly proved unexpected, as shown by the penalties and settlements imposed on Google over its location tracking on Android \citep{accc22, texasag25}. Previous measurement studies of location data have mostly focussed on Android and on third-party apps \citep{sivan19, girish25}. To our knowledge there is no previous systematic measurement study of the location data that iOS itself sends to Apple that decodes the message contents. Apple's public documentation does not describe network connections, endpoints or message contents.

Measuring what a phone sends requires understanding what the data means. The traffic can be decrypted on a jailbroken handset, but it is binary-encoded protobuf for which the message schemas are not public, and so they must be recovered from the software that sends it. Apple's code is mostly Objective-C (ObjC) compiled to ARM64. Decompilers give crude output, and in practice the analysis is manual inspection of assembly. Reverse engineering of this kind is skilled, labour intensive and slow, and it is a barrier to independent scrutiny of what phones send. Using AI to assist with it, or even to replace it, is therefore attractive. Whether this is viable matters for forensic and privacy analysis generally, and not only for the present study.

Our main contributions are:

\begin{enumerate}
\def\labelenumi{\arabic{enumi}.}
\tightlist
\item
  \emph{Direct AI reverse engineering is not viable.} We use two frontier models (Claude Opus 4.8, Codex GPT Sol 5.6) to reconstruct the schemas of protobufs observed in our traffic. The great majority of the schemas contain substantial errors, and these are hard to detect because the schemas are internally consistent. Direct AI use is therefore unsuitable for forensic and privacy analysis.
\item
  \emph{AI-written analysis tools are viable when they can be validated.} We use Claude to write a Python tool, \texttt{apple\_protobuf\_extractor} (``Extractor''), that recovers protobuf schemas from Apple's binaries. We can make this reliable because we are able to validate the output. Namely, we reverse engineer the template that Apple's protobuf classes follow, compile protobufs with known schemas to ARM64 with a compiler of the same era as the Apple code, and require Extractor to recover them exactly. We also check that (i) the recovered schemas decode our captured traffic to sensible values and (ii) they match previously published hand-decoded schemas. This validation needs reverse engineering and measurement of its own, so it is not free, but it is much less work than recovering the schemas by hand.
\item
  \emph{Measurements of the location data sent to Apple.} Using Extractor to decode the traffic of an iPhone 8 (iOS 16.7.11), we find that: (i) no message carries a direct identifier of the user or handset\footnote{Handset model, OS version, locale and carrier name are transmitted but these can be expected to be shared by many phones and so not be individually identifying.}, (ii) with location on, the handset sends its precise position to Apple in reverse geocoding and Wi-Fi fingerprint lookups which carry the entry and exit times of visits, in Wi-Fi and cell lookups, and in batched uploads of GPS fixes and Wi-Fi and cell observations, (iii) usage telemetry carries session identifiers that persist for at least 106 days and so link a handset's records over months, and (iv) with location off, the MAC address of the connected Wi-Fi access point, a network identifier, is still sent, which locates the handset to tens of metres if Apple knows where that access point is. Apple's documentation does not mention (iii) or (iv).
\end{enumerate}

Extractor (plus associated tool \texttt{ios\_decoder} and the extracted location services protobuf descriptors) are available on GitHub at https://github.com/doug-leith/ios-protobuf-decoding. Our AI analysis data (Section \ref{sec:AI}), including the prompts used, and our network measurement data (Section \ref{sec:network}) are available on GitHub at https://github.com/doug-leith/ios-location-services-study-data.

The rest of the paper is organised as follows. We review related work, then evaluate direct use of two LLMs, Claude and Codex, for reverse engineering and describe Extractor and its validation. We then apply Extractor to Apple's location services: we describe the experimental setup and the architecture of the services, and analyse the location data sent, the sensor data uploads, the telemetry and the identifiers used, before comparing our observations with Apple's public documentation. We end with the limitations of our measurements and our conclusions.

\section{Related Work}\label{related-work}

\emph{LLMs for reverse engineering}. LLMs have been applied to decompilation \citep{llm4decompile} and to recovering types and symbols from stripped binaries \citep{resym}, including Objective-C method signatures in macOS frameworks \citep{motif}. Benchmarks report that the best agents solve only 27\% to 32\% of realistic binary-analysis tasks \citep{srebench}, and that decompiled code can pass tests yet diverge in behaviour \citep{liu26}, consistent with our finding that errors are hard to spot.

\emph{Recovering protobuf schemas}. Existing tools recover schemas from binaries that embed a serialised descriptor \citep{pbtk, protodump}, or infer them from traces \citep{blackboxprotobuf}, and neither applies to Apple's \texttt{PBCodable} classes. We are not aware of prior work that recovers protobuf schemas from compiled \texttt{PBCodable} code and checks them against compiled ground truth.

\emph{Network traffic analysis}. A number of large-scale measurement studies have used network traffic analysis to detect transmissions to tracking services by Android apps, and to investigate the data transmitted by mobile apps. These use rooted phones and a man-in-the-middle attack to intercept and decrypt HTTPS traffic leaving the phone. Typically, apps are installed from the Google Play store, run for a short period with no user interaction, and the network traffic is recorded and analysed for transmission of sensitive information such as location and personal identifiers. See, for example,~\citep{274431, 274614, iphones, infocomgaen21, securecom21, snooping21, securecom22, narseo18, kollnig22}.

\emph{Mobile OS-level telemetry}. While there has been a fair number of measurement studies on app telemetry, mostly telemetry sent by third-party SDKs, there has been much less work on data transmitted by the mobile OS itself. The earliest work on Android appears to be~\citep{infocomgaen21} with follow-on studies by~\citep{securecom21, snooping21, securecom22}. The earliest work on Apple iphones appears to be~\citep{securecom21}, with more recent work focussing on the Apple App Store~\citep{apple_app_store_2026}.

\emph{Apple location services}. Apple's Wi-Fi positioning service has been reverse engineered from captured traffic and used to look up access point locations \citep{isniffgps}, and Rye and Levin show that it can be used to learn the locations of over 2 billion access points \citep{10646810}. A repository that cites that paper \citep{acheong08} adds protobuf field names recovered from \texttt{CoreLocationProtobuf}, tools for the tile and cell-tower lookups, and an informal technical report. The report records traffic from \texttt{locationd}, \texttt{geod} and \texttt{geoanalyticsd}, but most payloads are not decoded and it reaches conclusions about the data sent without recovering the message formats. We are not aware of a systematic measurement of what these daemons send that decodes the payloads, nor of one that covers \texttt{geod} and the CoreRoutine traffic. On-device forensic work documents what iOS stores about location \citep{edwards18, elcomsoft18} (the CoreRoutine databases, and retention of Significant Locations), which is the on-device counterpart of the uploads that we observe.

\emph{Anonymity of location data}. Removing identifiers does not make a location trace anonymous: a few coarse spatio-temporal points can single out most individuals \citep{demontjoye13, zang11, golle09}. We therefore report the session and network identifiers, as well as the location accuracy.

\section{LLM Reverse Engineering Is Too Lossy}\label{sec:AI}

\subsection{Evaluation Setup}\label{evaluation-setup}

We evaluate how accurately two frontier LLMs (Claude Opus 4.8, Codex GPT Sol 5.6) reconstruct Apple \texttt{PBCodable} protobuf descriptors directly from Mach-O disassembly.

Starting from one seed protobuf message class, the LLM is asked to recover the name, wire type, number and label (\texttt{optional}/\texttt{required}/\texttt{repeated}) for every field, and to recurse into every protobuf message- or enum-typed field until every reachable type has been recovered. An orchestrating Python script assigns recovery of each message- or enum-type to a fresh sub-agent with its own clean context window\footnote{Each message type corresponds to an ObjC class with its own \texttt{writeTo:}/\texttt{readFrom:} methods. Assigning a fresh agent to each class means each agent analyses a single pair of \texttt{writeTo:}/\texttt{readFrom:} methods.}. The prompt instructs the LLM to leave a field marked \texttt{//\ UNRESOLVED:\ \textless{}reason\textgreater{}} rather than guess when disassembly genuinely does not reveal something, and to keep a field as \texttt{bytes} if that is its genuine wire type rather than invent a message type for it. The script maintains a running record of message types already reconstructed, so each message- or enum-type is only reconstructed once even if it occurs multiple times within the seed message and its sub-messages.

The evaluation is blind. Each sealed LLM runs in a sandbox that denies it the ground-truth \texttt{.proto} files, any prior session transcript, and web access. It can use only the disassembly tool and read only the raw dyld shared cache. Grading happens afterwards, outside the sandbox, by diffing the LLM's reconstructed protobuf descriptor field-by-field against the schema produced by our independently validated Extractor tool, which the LLM never sees during the run.

\subsection{Results}\label{results}

As seed messages we use 17 GeoServices and CoreLocationProtobuf request and response protobufs observed in network traffic. Each seed is analysed separately, so a message or enum type that occurs in several seeds is recovered afresh in each, and its fields are counted once per seed. The 17 seeds contain 7,788 fields (4,987 distinct) in 1,627 message types. Of these fields, 4,151 have a scalar type (integer, boolean, floating point, string or bytes), 993 an enum type and 2,644 a message type.

We compare each LLM descriptor with Extractor's by following field numbers outward from the seed message, as a decoder does, rather than by matching type names. A field is correct if the LLM descriptor has a field with the same number at the same place in the message tree (that is, reached from the seed message by the same sequence of field numbers), with the same type and the same name. For a scalar field the type must be identical, including signedness and encoding (\texttt{int32}, \texttt{uint32}, \texttt{sint32}, \texttt{fixed32}), since these decode the same bytes to different values. For an enum- or message-typed field we report two levels: the field refers to the right enum or message, and, more strictly, every value of that enum, or every field of that message and of all messages below it, is also correct.

Where an LLM and Extractor disagree we checked the binary (the writer function each field's \texttt{writeTo:} calls, the instance variable (ivar) type encoding and the enum value-string tables) by a mixture of scripted and manual analysis. Of the 413 such disagreements in which the LLM gives a type, Extractor is found to be correct. We also picked 50 fields at random where Claude, Codex and Extractor agree and manually analysed them. In every case the manual analysis also agreed.

\begin{table}[tb]
\centering
\footnotesize
\caption{Fields recovered correctly (number, type and name). }\label{tab:accuracy}
\begin{tabular}{|l|r|r|r|}
\hline
\textbf{Field type} & \textbf{Fields} & \textbf{Claude} & \textbf{Codex} \\
\hline
Scalar & 4,151 & 3,575 (86.1\%) & 3,283 (79.1\%) \\
\hline
Enum: right enum & 993 & 807 (81.3\%) & 602 (60.6\%) \\
\quad all values correct & 993 & 775 (78.0\%) & 575 (57.9\%) \\
\hline
Message: right message & 2,644 & 2,337 (88.4\%) & 1,923 (72.7\%) \\
\quad whole subtree correct & 2,644 & 1,528 (57.8\%) & 1,374 (52.0\%) \\
\hline
\end{tabular}
\end{table}

\subsubsection{Analysis Of The Errors}\label{analysis-of-the-errors}

\begin{table}[tb]
\centering
\footnotesize
\caption{What a decoder does with the bytes of each field that is not correct (counts are fields, each counted once per seed it occurs in). Bold rows group the errors by consequence, with an example; the indented rows give the type of error behind each. Apart from descriptors that do not load, and the occasional message whose bytes no longer parse, every error decodes without complaint.}\label{tab:errors}
\begin{tabular}{|>{\raggedright\arraybackslash}p{5.85cm}|r|r|}
\hline
 & \textbf{Claude} & \textbf{Codex} \\
\hline
\textbf{Descriptor does not load:} it names a type it never defines\newline\textit{e.g.\ Codex gives \texttt{CoarseLocationMetadata.type} an enum type \texttt{Type} that it never defines} & \textbf{67} & \textbf{24} \\
\quad Enum type never defined & 52 & 24 \\
\quad Message type never defined & 13 & 0 \\
\quad Type of a scalar field never defined & 2 & 0 \\
\hline
\textbf{Value silently dropped, or left as raw bytes}\newline\textit{e.g.\ Claude types the latitude \texttt{LatLngE7.latE7} (\texttt{sfixed32}) as \texttt{int32}, so a decoder skips it} & \textbf{38} & \textbf{61} \\
\quad Message left as \texttt{bytes} & 29 & 1 \\
\quad Field omitted & 1 & 52 \\
\quad Incompatible wire type & 5 & 4 \\
\quad Enum and message confused & 3 & 4 \\
\hline
\textbf{Wrong value or wrong enum name}\newline\textit{e.g.\ Claude types \texttt{CdmaCellNeighbor.pnoffset} (\texttt{sint32}) as \texttt{int32}, so $-5$ decodes as 9} & \textbf{56} & \textbf{31} \\
\quad A different enum & 24 & 24 \\
\quad \texttt{sint} read as \texttt{int} & 15 & 0 \\
\quad A different message & 10 & 4 \\
\quad Same encoding, different meaning (e.g.\ \texttt{double} for \texttt{sfixed64}) & 3 & 1 \\
\quad Strings or bytes read as packed integers & 3 & 0 \\
\quad Integer read as an enum & 1 & 2 \\
\hline
\textbf{Wrong only when the top bit is set} (signed vs unsigned)\newline\textit{e.g.\ Claude types \texttt{ChargerPlug.maximumPower} (\texttt{uint32}) as \texttt{int32}, so 3,000,000,000 would decode as $-1{,}294{,}967{,}296$} & \textbf{128} & \textbf{3} \\
\hline
\textbf{Right value, label lost or wrong}\newline\textit{e.g.\ Codex types \texttt{MotionActivity.type} as \texttt{int32}, so a number is shown instead of the activity name} & \textbf{50} & \textbf{136} \\
\quad Enum read as a plain integer & 6 & 90 \\
\quad Wrong field name & 28 & 46 \\
\quad Placeholder field name & 16 & 0 \\
\hline
\textbf{Wrong at the field itself} (all of the above) & \textbf{339} & \textbf{255} \\
\hline
\textbf{Cut off by an earlier error:} never decoded, or decoded under the wrong schema\newline\textit{e.g.\ Codex swaps fields 1, 2 and 8 of \texttt{PDPlaceRequest}, so none of the 34 captured requests parses} & \textbf{730} & \textbf{1,725} \\
\quad Recovered but on the wrong field, correct there & 4 & 1,680 \\
\quad Recovered but on the wrong field, wrong there & 0 & 43 \\
\quad Never recovered & 726 & 2 \\
\hline
\textbf{Total not correct} (of 7,788 fields) & \textbf{1,069} & \textbf{1,980} \\
\hline
\end{tabular}
\end{table}

Table \ref{tab:errors} breaks down the observed errors in more detail. This shows that many of the errors are due to fields being ``cut off''. That is, an error higher up on the path from the seed message makes everything below it unreachable. A handful of errors is responsible for the cut-offs, e.g.~Claude's placeholder for \texttt{PDPlace.requestData} cuts off 697 fields, and Codex's wrong field numbers in \texttt{PDPlaceRequest} 1,725.

Table \ref{tab:errors} also shows that the two models fail in different ways.

\begin{itemize}
\tightlist
\item
  \emph{Claude.} Most of its errors at the field itself are integer signedness, mainly \texttt{int32} for \texttt{uint32}. These decode identically until a value reaches \(2^{31}\). It also leaves 29 message-typed fields as opaque \texttt{bytes}. For 52 enum fields the type name resolves to nothing: most are short, scope-relative names such as \texttt{Type} that are never defined; the rest are placeholder names.
\item
  \emph{Codex.} It types 90 enum fields as plain integers, which decode the number but lose its meaning, and omits 50 enum fields altogether.
\end{itemize}

Within enums that are otherwise identified correctly, Claude misnumbers 224 values in 11 enums and Codex 197 values in 7. For example, Claude omits value 35 (\texttt{DIRECTIONS}) of \texttt{DailyUseSummaryUseEventType} and shifts the 161 values after it down by one. Codex swaps values 3 and 12 of \texttt{MapRegionSourceType}. Every such value decodes from the wire to the wrong name.

A further 28 fields (Claude, plus 16 with placeholder names) and 46 (Codex) have the right type but the wrong name, e.g.~Codex names the two fields of \texttt{Attribution} that hold an attribution's name and URL \texttt{logo} and \texttt{logoChecksum}.

Neither model adds spurious field numbers to a message it identifies correctly. Both add types that are not part of the seed:

\begin{itemize}
\tightlist
\item
  \emph{Claude.} 24 types (115 fields). 22 of these (113 fields) are real GeoServices types reached through a wrong reference. For example, it substitutes the 30-field route-attributes message for the \texttt{WiFiESS.attributes} enum, together with the options messages below it, and the licence-plate region types (\texttt{GEOLPRRegion} and others, which it names \texttt{Region}, \texttt{RegionInfo} and so on) for the tile-region structure in \texttt{Resources}. One is an empty message standing in for an enum. One, a 2-field \texttt{TileType}, has no counterpart in the binary.
\item
  \emph{Codex.} Four types (8 fields). Three are real types and one is an empty message standing in for an enum.
\end{itemize}

Apart from Claude's 2-field \texttt{TileType} and the two empty stand-ins, every type the models add exists in the binary, but belongs elsewhere in the schema.

\subsubsection{Errors Tend To Concentrate}\label{errors-tend-to-concentrate}

\begin{figure}[tb]
\centering
\includegraphics[width=0.63\columnwidth]{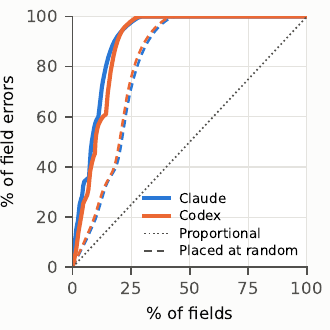}
\caption{Cumulative share of field-level errors against cumulative share of fields, with message types sorted descending by number of errors. Cut-off fields are excluded; the random curves are the mean of 2,000 random placements of the same errors. Half the errors lie in message types holding about a tenth of the fields, about twice as concentrated as random placement.}
\label{fig:field-error-concentration}
\end{figure}

The errors are concentrated in part of the schema, but not in a few messages. Figure \ref{fig:field-error-concentration} shows that half of them lie in message types holding about a tenth of the fields, roughly twice as concentrated as if the same number of errors had been placed at random. Even so, they touch some 200 message types, most of them with a single error. A single error makes every message above it incorrect, so these errors break far more subtrees than they touch directly.

\subsubsection{Variance Across Seed Protobufs}\label{variance-across-seed-protobufs}

\begin{figure*}[tb]
\centering
\includegraphics[width=0.78\textwidth]{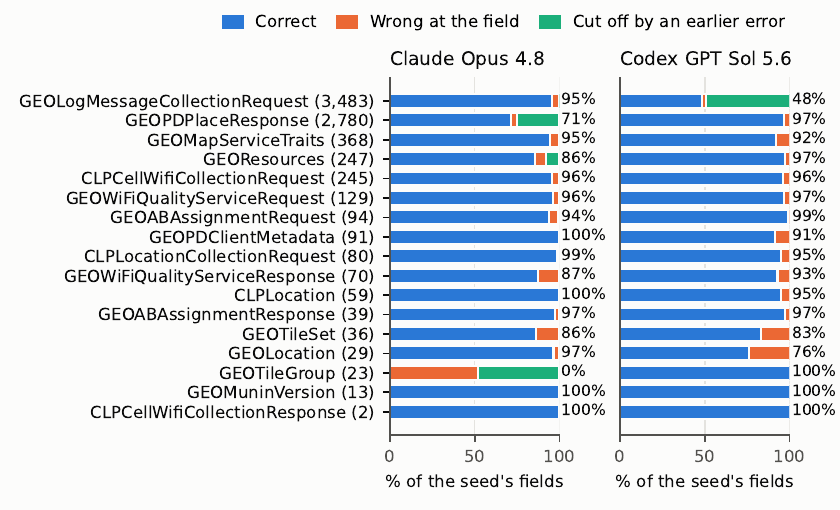}
\caption{Outcome for every field of each seed, as a percentage of the seed's fields (field count in brackets). 12 (Claude) and 14 (Codex) of the 17 seeds are at least 90\% correct; each model's losses are dominated by one error high in the message tree that cuts off a large subtree (Claude: \texttt{GEOPDPlaceResponse}; Codex: \texttt{GEOLogMessageCollectionRequest}).}
\label{fig:per-seed}
\end{figure*}

Figure \ref{fig:per-seed} shows the error rates for each of the seed protobufs. Each seed is analysed independently of the others, so this gives an indication of the level of variability that can be expected from using an LLM to reconstruct a protobuf. The reconstruction rates generally vary between about 85\% and 100\% but can be much worse, e.g.~Claude's \texttt{GEOTileGroup} is entirely wrong (Claude types all nine of the message's \texttt{uint32} fields as \texttt{int32} and replaces its three sub-message types with placeholders).

\subsubsection{Decoding Real Traffic}\label{decoding-real-traffic}

We also decoded real traffic with the recovered descriptors and compared the result with Extractor's decoding of the same bytes. We collected 1,057 instances of the seed messages from the iPhone 8 captures, including instances nested inside other messages, for 15 of the 17 seeds. The two exceptions:

\begin{itemize}
\tightlist
\item
  \texttt{GEOMapServiceTraits} is held only by on-device objects, such as the XPC request between an app and \texttt{geod}, so it never appears on the network.
\item
  \texttt{GEOMuninVersion} occurs only inside a part of \texttt{Resources} that none of the captured payloads contains.
\end{itemize}

Real traffic exercises only a small part of each schema. Of the 7,407 fields in these 15 seeds, only 806 (10.9\%) ever occur in the captures. On those 806 fields the recovered descriptors are more accurate than over the whole schema: 94.2\% correct for Claude and 95.9\% for Codex, against 85.8\% and 73.7\% over all fields of the same seeds. Errors therefore concentrate in rarely used fields, and testing a recovered descriptor against captured traffic would miss almost 90\% of the fields it gets wrong.

Where errors do fall on fields used in practice, the decoder reports nothing, but the decoded data is wrong:

\begin{itemize}
\tightlist
\item
  \emph{Lost data.} In Claude's descriptor for \texttt{GEOResources}, two fields refer to the wrong message type (\texttt{GEOResource.region} to the licence-plate \texttt{Region} message and \texttt{GEOTileGroup.tileSets} to \texttt{TileSet}). Decoding the eight captured \texttt{Resources} payloads silently loses 70\% of the decoded values and leaves a further 6\% as undecoded bytes.
\item
  \emph{Wrong values.} Claude types \texttt{CLPLocation.appBundleIdIndices}, a list of strings, as a list of integers, so the string \texttt{"0"} decodes as the number 48. Codex types the photo count \texttt{photosAvailable} as a boolean.
\item
  \emph{Parse failure.} The only error that announces itself is Codex's \texttt{PDPlaceRequest} field-number error, whose descriptor fails to parse any of the 34 captured messages.
\end{itemize}

\subsubsection{Compute Cost}\label{compute-cost}

\begin{figure}[htb]
\centering
\includegraphics[width=\columnwidth]{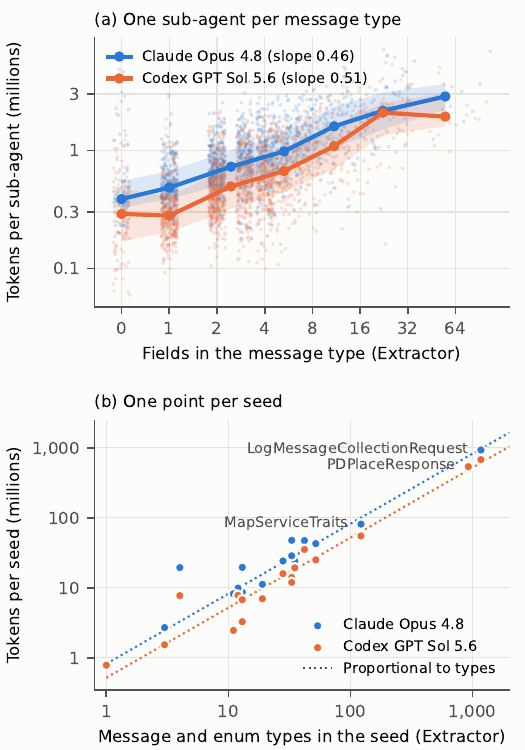}
\caption{Tokens used to recover descriptors. (a) One point per sub-agent that recovered a message type, against the number of fields Extractor gives that type; lines are medians within size bins and bands the interquartile range. Cost per sub-agent grows only with about the square root of the number of fields, from a large fixed cost. (b) One point per seed, against the number of message and enum types reachable from the seed; dotted lines are tokens proportional to the number of types. The cost of a seed is set by how many types it reaches.}
\label{fig:tokens-vs-size}
\end{figure}

As well as being error prone, using AI to recover protobuf descriptors is computationally expensive.
Recovering the 17 seed descriptors took 2,078 sub-agents for Claude and 2,333 for Codex, one for each message or enum type the model dispatched.
Figure \ref{fig:tokens-vs-size}(a) shows how the cost of one sub-agent depends on the size of the message type it recovers. There is a fixed cost of around 0.48 million tokens (Claude) and 0.28 million (Codex), plus a cost that grows with roughly the square root of the number of fields\footnote{The fitted slope on log--log axes is 0.46 for Claude and 0.51 for Codex, so doubling the number of fields raises the cost by about 40\%.}.

Figure \ref{fig:tokens-vs-size}(b) shows that total tokens per seed protobuf rise almost in proportion to the number of message and enum types in the seed (fitted slope 0.90 for Claude and 0.96 for Codex), at a median of 0.82 million tokens per type for Claude and 0.52 million for Codex. The two largest seeds, \texttt{GEOPDPlaceResponse} (925 reachable types) and \texttt{GEOLogMessageCollectionRequest} (1,172), account for 80\% of Claude's tokens and 85\% of Codex's.

Some seeds lie well above the dotted lines because a model dispatched more sub-agents than the seed has types: a wrong type reference pulls in types that do not belong to the seed, and each is recovered by its own sub-agent. For example, \texttt{GEOTileGroup} reaches four types. Codex dispatched four sub-agents for it. Claude dispatched 16, 15 of them for types outside the seed, reached through its three placeholder \texttt{group\_} types. Similarly, \texttt{GEOWiFiQualityServiceResponse} reaches 33 types. Claude's substitution of the route-attributes message for the \texttt{WiFiESS.attributes} enum pulls in that message and the 27 types below it, and its use of the \texttt{GEOTimeRange} class for an inline time-range structure adds one more, so 29 of Claude's 51 sub-agents for this seed recovered types outside it. Errors therefore raise cost as well as lowering accuracy.

We do not count 94 Claude sub-agent attempts (2.9 million tokens) and three Codex attempts that failed with infrastructure errors and were retried.

Although wall clock time is only an indirect measure of compute cost, to put the token counts in context, recovering the 17 seed protobufs took roughly five days of wall-clock time. In comparison, Extractor recovers the schemas of the whole GeoServices and CoreLocationProtobuf images (1,805 message types, which include the 17 seeds and every type they reach) in about 2 minutes (126 s and 4 s) on one core of an Apple M3 Max laptop, with no model calls. Its output is identical to the reference schemas used above.

\subsubsection{Choice Of Prompt}\label{choice-of-prompt}

The above results are for a recursive prompt, where each message type is decoded by its own agent. We also investigated decoding each seed protobuf using a single agent. However, we found this gave far worse performance. With the single-agent approach, only 696 of the 7,788 fields (8.9\%) are correct, mostly because the agent stops before reaching most of the nested messages (6,360 fields are never recovered), and only 59\% of the fields it does reach are correct. This may be due to the single agent exhausting its context window.

\subsection{Discussion}\label{discussion}

\subsubsection{Direct AI Analysis Is Too Unreliable}\label{direct-ai-analysis-is-too-unreliable}

This data indicates that direct protobuf reconstruction by Claude and Codex is far too unreliable for forensic and privacy analysis applications. Not only are the error rates too high, but the great majority of the errors cannot be easily caught since the protobuf descriptors generated are internally consistent: correct protobuf syntax, plausible field names, a type that compiles and decodes real messages. Nothing in the descriptor itself flags the error. In our captures only one error, Codex's field numbers in \texttt{PDPlaceRequest}, makes decoding fail; every other error decodes without complaint, and almost 90\% of the wrongly recovered fields never occur in the captured traffic at all. Catching these errors requires an external reference such as the ground-truth schema, or some other independent evidence.

\subsubsection{Fair Comparison}\label{fair-comparison}

Regarding independent evidence, Apple does not publish its ground-truth schemas. With hindsight, after development of Extractor, we can create our own ground-truth schemas and associated binaries. But these are a by-product of the detailed reverse engineering carried out while building Extractor, and so they are not ``free'': giving them to the direct-AI agents would hand them the tool's output and would not yield a fair comparison. Similarly, the network measurements reported later are another source of external evidence that could be fed back to the agents. However, these were not used during the development of Extractor, only for its subsequent validation. Additionally, as already noted above, their coverage of the protobuf fields, and so the feedback they could provide, is limited.

\section{Protobuf Extractor Tool}\label{sec:extractor}

While using AI to analyse binaries directly to extract protobuf schemas is too error prone, AI can nevertheless be used to develop a reliable Python tool to carry out this work. In this section
we describe such a tool, \texttt{apple\_protobuf\_extractor} (``Extractor''). Extractor is a 21,000-line Python tool, with a comparably sized test suite, built entirely by Claude. None of the code or tests is human-authored.

\subsection{From Descriptor To Assembly: A Worked Example}\label{from-descriptor-to-assembly-a-worked-example}

\begin{figure}[htb]
\centering
\begin{minipage}{0.62\columnwidth}
\begin{lstlisting}[style=protofig,basicstyle=\fontsize{5.9}{6.9}\selectfont\ttfamily]
message TestMessage {
  string first  = 5;
  int32  second = 2;
  float  third  = 9;
}
\end{lstlisting}
\end{minipage}
\caption{Example protobuf message, \texttt{TestMessage}.}
\label{fig:proto-example}
\end{figure}

\begin{figure}[htb]
\begin{lstlisting}[style=objcfig,basicstyle=\fontsize{5.9}{6.9}\selectfont\ttfamily]
@interface TestMessage : PBCodable {
    NSString *_first;
    int32_t   _second;
    float     _third;
    struct { unsigned has_first:1;
             unsigned has_second:1;
             unsigned has_third:1; } _has;
}
@end

@implementation TestMessage
- (void)writeTo:(PBDataWriter *)writer {
    if (_has.has_first)  PBWriteStringField(writer, _first, 5);
    if (_has.has_second) PBWriteInt32Field(writer, _second, 2);
    if (_has.has_third)  PBWriteFloatField(writer, _third, 9);
    [super writeTo:writer];
}
@end
\end{lstlisting}
\caption{ObjC runtime representation of \texttt{TestMessage}.}
\label{fig:objc-example}
\end{figure}

Consider the protobuf message in Figure \ref{fig:proto-example}. Figure \ref{fig:objc-example} shows how this would be implemented by Apple as an ObjC class\footnote{The ObjC implementation is based on a template extracted by reverse engineering analysis.}. This class has one
ivar per field, a \texttt{\_has} bitfield struct recording which optional fields are set, and a
\texttt{writeTo:} method that calls a field-writing helper, once per set field, passing the field number as a literal
argument. A production Apple binary carries only the compiled result of code such as
this. The compiled binary has no field names, no field numbers and no indication that a given ivar
even corresponds to a protobuf field. When compiled, the method above becomes the code shown in Figure \ref{fig:asm-example}.

\begin{figure}[htb]
\begin{lstlisting}[style=asmfig,basicstyle=\fontsize{5.9}{6.9}\selectfont\ttfamily]
    ldrsw  x21, [x8, #ivar_off(_has)]
    ldrb   w8,  [x20, x21]      ; w8 = _has bitfield byte
    tbnz   w8, #0, L_first      ; has_first?
    tbnz   w8, #1, L_second     ; has_second?
    tbz    w8, #2, L_epilogue   ; !has_third -> done

L_third:
    ldrsw  x8, [x8, #ivar_off(_third)]
    ldr    s0, [x20, x8]        ; s0 = self->_third (a float, not x1)
    mov    x0, x19              ; x0 = writer
    mov    w1, #0x9             ; w1 = field number
                                ; (float/double only: w1, not w2)
    bl     _PBWriteFloatField
    b      L_epilogue

L_first:
    ldrsw  x8, [x8, #ivar_off(_first)]
    ldr    x1, [x20, x8]        ; x1 = self->_first (an NSString*)
    mov    x0, x19
    mov    w2, #0x5
    bl     _PBWriteStringField
    ldrb   w8, [x20, x21]
    tbz    w8, #1, L_third      ; re-test has_second/has_third
    b      L_second

L_second:
    ldrsw  x8, [x8, #ivar_off(_second)]
    ldr    w1, [x20, x8]        ; w1 = self->_second (a plain int32)
    mov    x0, x19
    mov    w2, #0x2
    bl     _PBWriteInt32Field
    ldrb   w8, [x20, x21]
    tbnz   w8, #2, L_third
    b      L_epilogue
\end{lstlisting}
\caption{Compiled \texttt{-Os} ARM64 for the \texttt{writeTo:} method in Figure \ref{fig:objc-example} (prologue, epilogue and retain/release housekeeping omitted).}
\label{fig:asm-example}
\end{figure}

The recovery task is to invert this. From the raw instruction stream
alone, the tool must identify: (a) that \texttt{\_first}/\texttt{\_second}/\texttt{\_third} are
protobuf fields at all, distinguished from scratch or internal state by the
ivar-offset idiom (\texttt{ldrsw} from a named
\texttt{\_OBJC\_IVAR\_\$\_Class.\_ivar} symbol); (b) each field's \emph{name} (from which
ivar-offset feeds the load immediately before the call);
(c) each field's \emph{number} (from the \texttt{mov\ w1,\#N}/\texttt{mov\ w2,\#N} immediately
preceding the \texttt{bl}, not from source or declaration order: fields 5, 2, 9
above are deliberately non-monotonic, to show that a tool that used ivar-declaration-order numbering would
fail);
(d) each field's \emph{type} (from which \texttt{PBWrite*Field} helper is called).

Extractor implements this by walking the method's
control-flow graph from entry, recording every \texttt{PBWrite*Field}-family call
site reached, and binding, for each one, a field number to the ivar name
that fed the value load.

\subsection{Complications Beyond The Simple Case}\label{complications-beyond-the-simple-case}

\subsubsection{A Full Control-Flow-Graph Walk Is Needed}\label{sec:cfg-walk}

The example above already shows why a textual scan is insufficient: the
compiler does not lay the three write blocks out in source order, and does
not reach a given block only by falling straight through to it. A tool scanning top-to-bottom for
\texttt{mov\ w1,\#N}/\texttt{mov\ w2,\#N;\ bl\ PBWrite*Field} would still find all three calls
in a function this small, but in physical-address order (9, 5, 2), with no
principled way to confirm each call is genuinely reachable, or to exclude
a similarly-shaped but dead call left over elsewhere in the binary.
Extractor therefore performs an explicit reachability walk over the
method's basic blocks from its single entry point, so every recovered
field write is guaranteed reachable, and later analysis stages can reuse
the same walk to resolve values that depend on which block a load
appears in, not merely that it appears somewhere in the function body.

\subsubsection{Repeated (Message-Typed) Fields}\label{sec:repeated-fields}

A \texttt{repeated\ Msg\ field\ =\ N} field is stored as a plain, untyped array. Nothing about the array itself records what kind of message its elements
are. A class states this fact in one of two other places, and the tool
must find whichever one a given class actually uses:

\begin{enumerate}
\def\labelenumi{\arabic{enumi}.}
\tightlist
\item
  \textbf{A small helper method the class defines purely to advertise its
  element type}. This is the easy case: the tool just reads off what class
  the method returns.
\item
  \textbf{\texttt{readFrom:}}: the element type has to be inferred from
  \texttt{readFrom:}, the method that parses an incoming byte stream back into
  an object (the reverse of \texttt{writeTo:}). \texttt{readFrom:} routes each
  incoming field to per-field handler code, and the tool follows that
  same routing to see which other class's parsing method a given
  field's handler calls.
\end{enumerate}

The second, harder route might seem the natural one to look for, but it
is the less common one in practice.
A third, rarer route resolves the element type by tracing which of the
class's own array-adding methods the routing logic calls directly. It is used
when neither of the above applies.

\subsubsection{Distinguishing Plain Ints From Sint/Fixed Variants}\label{sec:int-variants}

An ObjC int might be encoded on the wire as an \texttt{int32}/\texttt{int64}, a
zigzag-encoded \texttt{sint32}/\texttt{sint64}, or a fixed-width \texttt{fixed32}/\texttt{fixed64}. ObjC storage
does not distinguish these, so the tool works it out from the write call itself, but the details
depend on where the binary lives.

\texttt{PBDataWriter}, the class that builds the wire bytes, gives each
wire-type family its own method, e.g.~\texttt{writeInt32:},
\texttt{writeFixed32:}. In a standalone binary, \texttt{writeTo:}
calls these as Objective-C messages, so the method name says the
type outright: no disassembly is needed, just read the selector. Inside the
dyld shared cache (where most Apple frameworks reside), the same call is
implemented as an unnamed C function call. In this case, the tool follows the
call and reads the first few instructions of whatever it finds to decide on the wire type, e.g.~a
zigzag encoder shifts the value left by one bit before anything else.

\subsubsection{Enum Values: Another Lookup, In One Of Several Shapes}\label{sec:enum-values}

An enum field's value names, like the field names themselves, are not
stored as data. Many classes have a small companion method, named after
the field itself (e.g.~a \texttt{status} field's values are named by a
\texttt{statusAsString:} method), that takes the stored integer and returns the
matching name string. In this case, the tool recovers the field's full number-to-name
table by decoding that method. The compiler lays
this lookup out differently depending on how the values are distributed,
so the tool needs a different decoder for each shape:

\begin{enumerate}
\def\labelenumi{\arabic{enumi}.}
\tightlist
\item
  \textbf{Densely packed values} (e.g.~0 to 7): the compiler builds a small
  jump table, one entry per value, each pointing at a tiny block of code
  that returns that value's name. The tool walks the table block by
  block.
\item
  \textbf{Widely spaced values} (e.g.~20, 30): rather than a
  table, the compiler writes a sequence of individual ``is the value 20?
  is it 30? \ldots{}'' checks. The tool finds and reads each check in
  turn.
\item
  \textbf{Both at once}: a single field can use both shapes for different
  parts of its range, so full recovery needs both decoders to run on the same field and combine
  their results.
\item
  \textbf{No companion method}: some fields have no dedicated method
  like \texttt{statusAsString:}. Fortunately, Apple's runtime also gives every class a
  separate debugging method that builds a dictionary of every field's
  current value, by name, and so the tool reads the enum names there instead. The dense/sparse split
  above still applies.
\end{enumerate}

\subsection{Program Architecture}\label{program-architecture}

The codebase splits into two layers. \texttt{macho/} is a general-purpose Mach-O
and ARM64 binary-analysis layer with no protobuf-specific knowledge:
\texttt{memory.py} maps a loaded image (a standalone binary or a slice of the
dyld shared cache) and resolves chained-fixup pointers;\footnote{A binary's real load address is not known until
  runtime, so a stored pointer that may need to address something outside
  the binary itself (another loaded library, heap-allocated memory) has to
  be patched to the correct address once loading happens. Chained fixups
  are Apple's current format for that patching.}
\texttt{arm64decode.py}
decodes individual raw instruction words (\texttt{movz}, \texttt{adrp}, \texttt{cmp}, \texttt{ldr},
conditional branches, and so on); \texttt{disasm.py} iterates an instruction
stream; \texttt{classlist.py} and \texttt{strings.py} enumerate ObjC classes, ivars,
methods and string tables from a binary's metadata sections; \texttt{dyld\_cache.py},
\texttt{local\_symbols.py} and \texttt{cache\_classlist.py} handle shared-cache-specific
image lookup and symbolication, used only for cache-native runs.

\texttt{proto/} is the protobuf-recovery layer, built on top of \texttt{macho/}, summarised in Table \ref{tab:proto-modules}.

\begin{table*}[htb]
\centering
\footnotesize
\caption{\texttt{proto/} module responsibilities.}\label{tab:proto-modules}
\begin{tabular}{|p{4.0cm}|p{12.5cm}|}
\hline
\textbf{Module} & \textbf{Responsibility} \\
\hline
\texttt{protocore.py} & ivar type encoding to proto scalar/message type and label \\
\hline
\texttt{writeto.py} & \texttt{writeTo:} CFG walk, field number/name binding (Section \ref{sec:cfg-walk}) \\
\hline
\texttt{dispatch.py} & \texttt{readFrom:} field-dispatch decoding, used for repeated-element typing (Section \ref{sec:repeated-fields}) \\
\hline
\texttt{repeated.py} & repeated-field element type/ivar resolution (Section \ref{sec:repeated-fields}) \\
\hline
\texttt{enums.py} & both enum-recovery mechanisms across dense/sparse/hybrid shapes (Section \ref{sec:enum-values}) \\
\hline
\texttt{write\_helpers.py} & classifies a \texttt{writeTo:} helper's wire-type family from its tag-building instructions (Section \ref{sec:int-variants}) \\
\hline
\texttt{extractor.py} & orchestrator: walks the classlist, invokes the modules above per class, assembles \texttt{ProtoMessage}/\texttt{ProtoField} records and assigns each field number's confidence tier \\
\hline
\texttt{models.py} & plain data classes (\texttt{ProtoField}, \texttt{ProtoMessage}) shared across modules \\
\hline
\texttt{emitter.py} & renders recovered records as \texttt{.proto} text \\
\hline
\texttt{config.py}, \texttt{defaults.py} & class name to proto package/prefix mapping \\
\hline
\texttt{enums\_debug.py} & opt-in diagnostic sink (\texttt{--debug-enums}), dormant unless explicitly enabled, recording why a field's enum recovery failed \\
\hline
\texttt{protoclass\_heuristic.py} & a retired heuristic class detector, kept only as a fail-loud safety net if a class is ever routed through it instead of the current structural detector \\
\hline
\end{tabular}
\end{table*}

\subsection{Code Development}\label{code-development}

The full AI session transcripts, the complete commit history and AI-authored process artefacts (a running to-do list, a heuristics registry, its own limitations note) generated during the development of Extractor are available at https://github.com/doug-leith/extractor-code-development. The development process, including the AI errors made during it and the reliability of the AI's claims about its own work, is analysed in \citep{leith_code_development_2026}.

\subsection{Testing}\label{testing}

\subsubsection{Using Ground Truth From A Reverse-Engineered Template}\label{using-ground-truth-from-a-reverse-engineered-template}

Apple protobuf schemas are not publicly available, and so we cannot verify Extractor
output by comparison against them. Nevertheless, we can still
validate Extractor's output as follows.

Apple's ObjC protobuf implementation uses a separate class for each protobuf message and follows a fixed template as illustrated in Figure \ref{fig:objc-example}, repeated across every class. Having reverse-engineered that template closely enough to
reproduce it, we can author our own protobuf descriptors with a known schema,
write the equivalent ObjC, compile it to an ARM64 Mach-O binary, and run Extractor on it. Because the
schema was authored by us, any mismatch between the recovered
output and the authored ground truth is a tool defect. By testing against a variety of protobuf descriptors we can exercise Extractor's code thoroughly (see below). Descriptors are chosen to (i) trigger a specific code shape, facilitating testing, e.g.~an 8-field message compiles to a jump-table dispatch, a smaller one compiles to a compare chain, and (ii) reproduce Apple's field and enum names and values.

The compiler used also affects the code shapes observed in the ground-truth tests. Using Apple clang 11.0.0 (build clang-1100.0.33.8, bundled with Xcode 11.1) with the \texttt{-Os} optimisation flag reproduces the branch and jump-table shapes seen in the iPhone 8 binaries. For the more recent OS versions Apple clang 15.0.0 (clang-1500.3.9.4, Xcode 15.4) reproduces the patterns seen in Apple binaries\footnote{Newer compiler optimisations introduce some new binary code patterns not seen in the older compiler, as well as support for ARM64e and PAC authenticated pointers.}.

The Extractor test suite therefore consists of two kinds of tests. The first are ground-truth tests against binaries derived from known protobufs, and the second are regression tests against Apple binaries. The latter are regression tests because the true protobuf schemas are unknown, and so the tests can only confirm the tool's output has not changed, not that it was correct to begin with.

Table \ref{tab:coverage} shows the statement and branch coverage of the ground-truth tests. Note that much of Extractor's codebase by line count is either standard disassembler components not specific to the task, or plumbing that the synthetic fixtures are not meant to exercise. The data in Table \ref{tab:coverage} is therefore restricted to: (i) the modules that implement the protobuf-specific
recovery work, and (ii) within each module to the functions that actually decode instructions or Mach-O structures into a recovered name, number, type or table entry (excluding validation,
defensive and orchestration code). The code doing the actual binary analysis is well exercised by the ground-truth tests, with overall statement coverage of 83\% and branch coverage of 71\%.

\begin{table*}[htb]
\centering
\footnotesize
\caption{Statement and branch coverage of the specialised recovery code.}\label{tab:coverage}
\begin{tabular}{|p{3.0cm}|p{7.4cm}|p{2.4cm}|p{2.4cm}|}
\hline
\textbf{Module} & \textbf{Recovers} & \textbf{Statement coverage} & \textbf{Branch coverage} \\
\hline
\texttt{protocore.py} & ivar type/label decoding & 89.2\% (33/37) & 87.5\% (21/24) \\
\hline
\texttt{dispatch.py} & \texttt{readFrom:} field-dispatch decoding for repeated-element typing (Section \ref{sec:repeated-fields}) & 88.4\% (274/310) & 78.5\% (135/172) \\
\hline
\texttt{writeto.py} & \texttt{writeTo:} field number/name binding (Section \ref{sec:cfg-walk}) & 84.0\% (247/294) & 70.1\% (143/204) \\
\hline
\texttt{write\_helpers.py} & wire-type family classification (Section \ref{sec:int-variants}) & 83.0\% (44/53) & 73.3\% (22/30) \\
\hline
\texttt{enums.py} & enum value-table recovery (Section \ref{sec:enum-values}) & 80.2\% (482/601) & 67.0\% (233/348) \\
\hline
\texttt{repeated.py} & repeated-field element resolution (Section \ref{sec:repeated-fields}) & 77.9\% (127/163) & 67.6\% (69/102) \\
\hline
 & & \textbf{82.8\% (1207/1458)} & \textbf{70.8\% (623/880)} \\
\hline
\end{tabular}
\end{table*}

\subsubsection{Using Manually Reverse-Engineered Descriptors}\label{using-manually-reverse-engineered-descriptors}

In addition, Extractor's output is compared against Apple protobuf descriptors from iCloud, iMessage and iOS location services already manually reverse engineered and published by other researchers (see Section \ref{sec:acheong-compare} in the Additional Material for details). In summary, the field numbers agree wherever both define a field, and in every case where the published descriptor disagrees with Extractor on a field's type, meaning or structure, the iPhone's binaries agree with Extractor.

\subsubsection{Using Network Measurements}\label{using-network-measurements}

We take two more steps to check Extractor's accuracy and robustness.

Firstly, Extractor is coded defensively, configured to fail loudly and to always flag uncertainty to the user, e.g.~if it returns a field's type as \texttt{bytes} because it is unsure about its real type. For all of the 889 (message, field) pairs observed being transmitted over the network (see below), Extractor flags no uncertainty\footnote{Extractor flags four kinds of uncertainty: estimated numbers, unresolved or guessed types, undetermined element types, and unrecovered map types.}, 15 fields are unknown but none is an Extractor miss (12 are newer fields declared in iOS 27, one is a field of an opaque pass-through class, and two are fields in server responses that neither the iPhone 8 nor the iOS 27 binaries declare).

Secondly, the extracted protobuf descriptors are used to decode serialised Apple protobufs observed being transmitted over the network, and we confirm that (i) decoding succeeds, which confirms that the descriptor field wire types are compatible with the observed protobuf,
and (ii) the decoded field values are consistent with the extracted field names, e.g.~a latitude field has a reasonable value given a phone's known location. Of the 889 (message, field) pairs observed, 297 are numeric or string fields confirmed to have values consistent with the field name, 98 are enums with values consistent with the expected range, 76 are booleans (trivially consistent), 313 are sub-message containers, and 105 have no applicable check.

\subsubsection{iOS And macOS Versions Tested}\label{ios-and-macos-versions-tested}

Extractor has been tested on dyld shared caches from an iPhone 8 (iOS 16.7.11; cache UUID f203ff1d-3ce6-3f7a-9849-791b1c1f4b8b), macOS 14.7 (UUID 8047f785-c408-34e1-a0a8-38ee34ab89f4), and the recent versions iOS 27.0 (UUID 9416075f-d579-30bc-a976-2b2b9a39270e) and macOS 27.0 (UUID ea2c265e-297c-39c2-8646-7d8a2dff648a).

\subsection{Other Protobuf Implementations}\label{other-protobuf-implementations}

In addition to Apple's ObjC \texttt{PBCodable} implementations, Extractor recovers schemas for swift-protobuf (including Apple's \texttt{InternalSwiftProtobuf} and its newer packed name-map format) and for the full C++ libprotobuf runtime, and detects, but does not decode, Google's ObjC \texttt{GPBMessage}, C++ libprotobuf-lite and protobuf-c.~All of the messages in the location services network traffic we decode are \texttt{PBCodable}.

\section{Applying Extractor To Apple Location Services}\label{sec:network}

\subsection{Experimental Setup}\label{experimental-setup}

\subsubsection{Hardware And Software Used}\label{hardware-and-software-used}

We use an Apple iPhone 8 running iOS 16.7.11\footnote{We used a pre-A12 iPhone as an unpatchable bootloader exploit allows jailbreaking for later versions of iOS than on more recent handsets. iOS 16.7 is, however, the latest version available since Apple has now stopped updates for pre-A12 handsets. See \texttt{https://theapplewiki.com/wiki/Jailbreak}.}, jailbroken using Palera1n v2.2.1\footnote{\texttt{https://github.com/palera1n/palera1n}} and with the Ellekit\footnote{\texttt{https://github.com/tealbathingsuit/ellekit}} Objective-C function hooker. We intercept network connections using \texttt{mitmdump} 13.0.0.dev. We bypass SSL certificate pinning within handset processes using a modified version of SSL KillSwitch\footnote{Available at \url{https://github.com/doug-leith/poor-privacy-practices-of-the-apple-app-store/tree/main/ssl-kill-switch3-DL}}.

\subsubsection{Test Protocol}\label{test-protocol}

The test protocol used is similar to that in \citep{apple_app_store_2026}. In summary, we carry out experiments with a privacy-conscious but busy/non-technical user, who when asked does not select options that share data with Apple but otherwise leaves handset settings at their default value. We use a two-step process to jailbreak the phone and allow decryption of HTTPS traffic. Firstly, we (i) jailbreak the phone using Palera1in with the Sileo package manager, (ii) use Sileo to download and install Ellekit and openssh, (iii) connect to the handset via ssh and install the Objective-C injection script used to bypass SSL certificate pinning. Then we insert a SIM with no credit (data access is disabled), factory reset the phone and the user goes through the onboarding process without using a network connection, selecting the settings shown in Table \ref{tab:settings} (Additional Material). We jailbreak the phone again using Palera1in. Ellekit, openssh and the SSL bypass script are all left installed by the factory reset and become activated again after this second jailbreak without the need for a network connection. The user connects to the test Wi-Fi network and installs the mitmproxy CA cert as a trusted certificate by downloading it from the special URL http://mitm.it supported by \texttt{mitmdump}, enabling it in the phone settings and setting it to be trusted. At this point system HTTPS network connections can now be intercepted and successfully decrypted, and we record all handset network traffic.

The rest of the experiment is divided into three parts. In the first part the user is not logged in to an Apple account and the handset location setting is off. In the second part the handset location setting is turned on and the user logs in to an Apple account. In the third part, the phone SIM is replaced with one that has credit (so data connectivity is active), the Apple App Store is used to install the OpenVPN app and the handset settings are configured to send all network traffic via the VPN. The handset is then taken on four car commuter journeys, namely morning and evening for two days. The Apple Maps app is briefly opened to confirm the reported handset location is accurate, and then closed.

\subsubsection{Decrypting HTTPS Connections}\label{sec:mitm}

In the first two parts of the experiment we route handset traffic via a Wi-Fi access point (AP) that we control, configure this AP to use \texttt{mitmdump} as a proxy \citep{mitmproxy} and adjust the firewall settings to redirect all Wi-Fi HTTP/HTTPS traffic to \texttt{mitmdump} so that the proxying is transparent to the handset. When a process running on the handset starts a new network connection, the \texttt{mitmdump} proxy pretends to be the destination server and presents a fake certificate for the target server. This allows \texttt{mitmdump} to decrypt the traffic. It then creates an onward connection to the actual target server and acts as an intermediary, relaying requests and their replies between the handset process and the target server while logging the traffic. In the third part of the experiment, handset traffic is routed via a VPN server which is similarly configured to use \texttt{mitmdump}.

System processes typically carry out checks on the authenticity of server certificates received when starting a new connection and abort the connection when these checks fail. This happens even when the mitmproxy CA cert is installed on the handset as a trusted certificate. To bypass these checks, the iPhone is jailbroken, Ellekit is installed and a custom Objective-C injection script is used to bypass SSL certificate pinning within handset processes.

\subsubsection{\texorpdfstring{\texttt{ios\_decoder}}{ios\_decoder}}\label{ios_decoder}

Extractor recovers the protobuf descriptors from Apple binaries. We developed a second tool \texttt{ios\_decoder} that applies them to network captures: it reads a saved mitmproxy capture offline and for each connection prints the decoded HTTP request and response.

In more detail, within \texttt{ios\_decoder} a table maps each connection, identified by host and URL path, to its format. For a protobuf payload the entry gives the message name\footnote{The protobuf message name is found by inspection of the appropriate binary: search for the \texttt{NSURLSession} instance associated with the URL (or connection name from \texttt{geo\_manifest}), then trace back to find the protobuf class from which the POST body is derived.}, and the descriptor recovered by Extractor is used to decode the body with the standard protobuf library. The entry also gives the framing, e.g.~the length-prefixed ARPC envelope that wraps many Apple requests. Other payloads are decoded by format: property lists, JSON, map tiles and so on. HTTP headers and URL query parameters are decoded in the same way, with the values the table identifies as coordinates or tile numbers converted to the area they denote.

Of the 1,500 \texttt{geod} and \texttt{locationd} connections we observe, 42\% carry protobuf bodies. These include all of the location uploads, the telemetry, A/B experiment and dispatcher exchanges, \texttt{wloc} lookups and the Wi-Fi quality lookups and tiles. The remainder are mainly tile and resource downloads.

\subsubsection{Measurement Data \& Additional Material}\label{measurement-data-additional-material}

Tables \ref{tab:geod-connections} and \ref{tab:locationd-connections} (Additional Material) detail the network connections that we observe being made by \texttt{geod} and \texttt{locationd}, together with their estimated purpose and a summary of their content.

\subsection{Architecture Of Apple Location Services}\label{architecture-of-apple-location-services}

There is no public Apple documentation on the internal operation of its location services, and only very limited prior reverse engineering work has been published. The following is based on analysis of the \texttt{geod}, \texttt{locationd} and \texttt{GeoServices} binary code (class and string names, selective disassembly of stripped binaries) combined with decoding of observed network transmissions (the network traffic analysis is covered in detail later and only summarised here).

\begin{figure}[tb]
\centering
\includegraphics[width=\columnwidth]{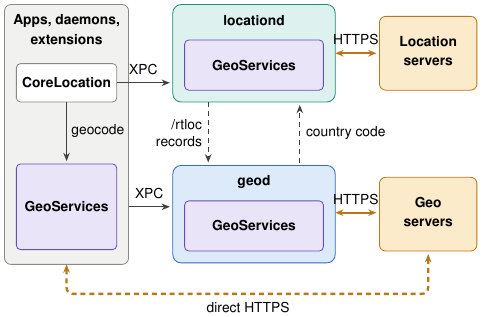}
\caption{Architecture of iOS location services. GeoServices (purple) is a single private framework loaded into \texttt{locationd}, \texttt{geod} and every process that uses it. Processes reach \texttt{locationd} through CoreLocation and \texttt{geod} through GeoServices, which MapKit and CoreLocation's geocoder also use. Thin arrows are XPC, thick arrows are HTTPS, and dashed black arrows are hand-offs between the daemons. In our captures two processes bypass \texttt{geod} and connect to the geo servers themselves (dashed amber arrow).}\label{fig:architecture}
\end{figure}

Location services on iOS are centred on two system daemons, \texttt{geod} and \texttt{locationd}, which make almost all the network connections analysed in this paper, and \texttt{GeoServices}, a large private framework (about 27 MB of code and 2,597 Objective-C classes) that both daemons link to. See Figure \ref{fig:architecture}.

\texttt{geod} is accessed via XPC and acts as a caching gateway between clients and Apple's geo servers. It serves map tiles, other map resources (stylesheets and icons), place data and place images from its caches where it can, and makes network requests on the client's behalf where it cannot. Other \texttt{geod} services that can make network requests include reverse geocoding, point-of-interest lookup, routing directions, ETA and transit updates and Wi-Fi quality lookups. \texttt{geod} provides a logging funnel, batching and uploading usage and log messages that clients report to it. It also maintains shared local state and logic, including: GeoServices configuration, country code (determined from the SIM or a geo-IP lookup), the A/B experiment assignment, request counters/throttling and offline map subscriptions. \texttt{geod} contacts some Apple servers without a client request, including the geo\_manifest server, and can preload tiles.

Only a few network endpoints used by \texttt{geod} are hard-coded in \texttt{GeoServices}. These include: (i) the defaults plist URL \texttt{configuration.ls.apple.com/config/defaults}, (ii) the geo\_manifest URL\footnote{\texttt{gspe35-ssl.ls.apple.com/}\allowbreak\texttt{geo\_manifest/}\allowbreak\texttt{dynamic/}\allowbreak\texttt{config\%\{DEVICE\_QUERY\}}} and (iii) the geo-IP country lookup URL \texttt{gspe1-ssl.ls.apple.com/pep/gcc}. The others are supplied by the responses from the geo\_manifest and defaults endpoints.

\texttt{locationd} hosts the phone location providers (GNSS, Wi-Fi, cell, BLE) and is accessed by apps via the public CoreLocation framework, which communicates with \texttt{locationd} over XPC. It also performs a range of other on-device processing, including geofencing, significant-location and visit monitoring, motion classification, health and fitness metrics (e.g.~VO\textsubscript{2} max, gait, fall detection), and Find My tracking avoidance. It downloads data from Apple for its own use, including: (i) tiles of cell identifiers and Wi-Fi MAC addresses with their locations from \texttt{cl2.apple.com}, presumably so that the phone location can be estimated on the handset from the cells and access points it observes, (ii) elevation and radio signal environment tiles from \texttt{cl3.apple.com} and (iii) GNSS assistance data (satellite orbit predictions) from \texttt{iphone-ld.apple.com}, which it sends to the GPS chipset. It can also transmit a Wi-Fi MAC address, cell identifier, or the handset location to Apple server \texttt{gs-loc.apple.com/clls/wloc}, which responds with the locations of nearby Wi-Fi networks and cell towers. \texttt{locationd} also reports (``harvests'') phone sensor data to Apple servers, either itself or, for the real-time traffic probe, through \texttt{geod} (Section \ref{sec:harvest}).

The clients of \texttt{geod} are system daemons, services and extensions that run without user action. They do not contact Apple's geo servers themselves: their GeoServices code passes requests to \texttt{geod} over XPC, and \texttt{geod} sends them to Apple. The clients are: \texttt{routined}, ``a per-user daemon that learns historical location patterns of a user and predicts future visits to locations''\footnote{Quoted from the macOS 14.7 manual page \texttt{routined(8)}. iOS has no manual pages, so the descriptions of \texttt{analyticsd}, \texttt{networkserviceproxy}, \texttt{photoanalysisd} and \texttt{mapspushd} below are taken from their macOS pages. The other clients have none.}, which apparently takes its locations from \texttt{locationd} and reaches \texttt{geod} through its XPC helper service \texttt{com.apple.CoreRoutine.helperservice} (``CoreRoutine'' for short); the Weather widget; \texttt{analyticsd}, which ``collects diagnostics and usage data locally for users that have opted in''; \texttt{networkserviceproxy}, a transparent network proxy for Apple system services; \texttt{kbd}, the keyboard daemon; the CoreLocation permission-prompt plugins; and \texttt{com.apple.wifi.ThreeBarsXPCService}, a prefetch and cache service for Wi-Fi network quality data. We infer the purpose of the clients without manual pages from their binaries.

Two further processes are not \texttt{geod} clients for network requests: they connect to Apple's geo servers directly. \texttt{photoanalysisd}, the photo library analysis agent (it supports Memories, People and photo search), sends reverse geocoding and point-of-interest lookups. Each carries a precise location, which in our captures is the handset position. The request code is in the MediaMiningKit framework, which links GeoServices, and the server URLs come from \texttt{geod}'s \texttt{geo\_manifest}. \texttt{photoanalysisd} reports its requests to \texttt{geod}'s telemetry. \texttt{mapspushd}, the Maps application services daemon, fetches Maps announcements and sends no location. It is a client of \texttt{locationd}.

\subsection{Analysis Of Location Data Sent}\label{sec:location_sent}

The following location data is observed being sent to Apple by \texttt{geod}, \texttt{locationd} and \texttt{photoanalysisd}:

\begin{enumerate}
\def\labelenumi{\arabic{enumi}.}
\tightlist
\item
  \emph{Coarse location is sent by both \texttt{geod} and \texttt{locationd} when location is on.} Tile downloads identify the area around the handset to within about 1 to 5 km for the finest tiles.
\item
  \emph{Precise location is sent by \texttt{geod}, \texttt{locationd} and \texttt{photoanalysisd} when location is on.} \texttt{locationd} sends Wi-Fi MAC addresses and cell identifiers for lookup and uploads GPS locations accurate to about 5 m. \texttt{geod} sends the real-time traffic probe (see below) and, on behalf of CoreRoutine, lookups that carry a precise location. \texttt{photoanalysisd} sends precise locations directly to Apple's geo servers.
\item
  \emph{Uploads are delayed.} Apart from the real-time traffic probe, which is sent within a minute, uploads are sent hours after the data was collected, and up to about 2 days after.
\item
  \emph{Switching location off removes most location data but not the MAC address of the connected access point.} The Wi-Fi quality service sends it even when location is off, which locates the handset to tens of metres if Apple holds the position of the access point. The location that \texttt{geod} sends for its other clients varies (see Table \ref{tab:geod-clients}).
\end{enumerate}

\subsubsection{Coarse Location}\label{coarse-location}

Both \texttt{geod} and \texttt{locationd} download tiles of map, Wi-Fi, cell and other data, and each request identifies an area. \texttt{geod} names a tile by its X, Y and zoom Z values in the standard Web Mercator scheme, which divides the Earth into \(2^Z \times 2^Z\) tiles. A tile is about 10 km wide at zoom 12, 5 km at zoom 13 and 1 km at zoom 15 at the equator; about 60\% of that at the latitude of Ireland.

\texttt{locationd} instead puts the latitude and longitude of the tile's south-west corner in the URL path. In \nolinkurl{cl2.apple.com/1/v1/143/173/1433500_1738000.gz}, the value 1433500 equals (latitude+90)*10000 and 1738000 equals (longitude+180)*10000, giving latitude 53.35\(^\circ\) and longitude \(-6.20^\circ\). Its Wi-Fi and signal environment tiles are about 5.6 km by 3.3 km in size, its elevation tiles about 11 km by 7 km, and its GSM and LTE cell tiles about 220 km by 130 km.

\begin{figure}[tb]
\centering
\begin{subfigure}[b]{0.45\columnwidth}
\includegraphics[width=\columnwidth]{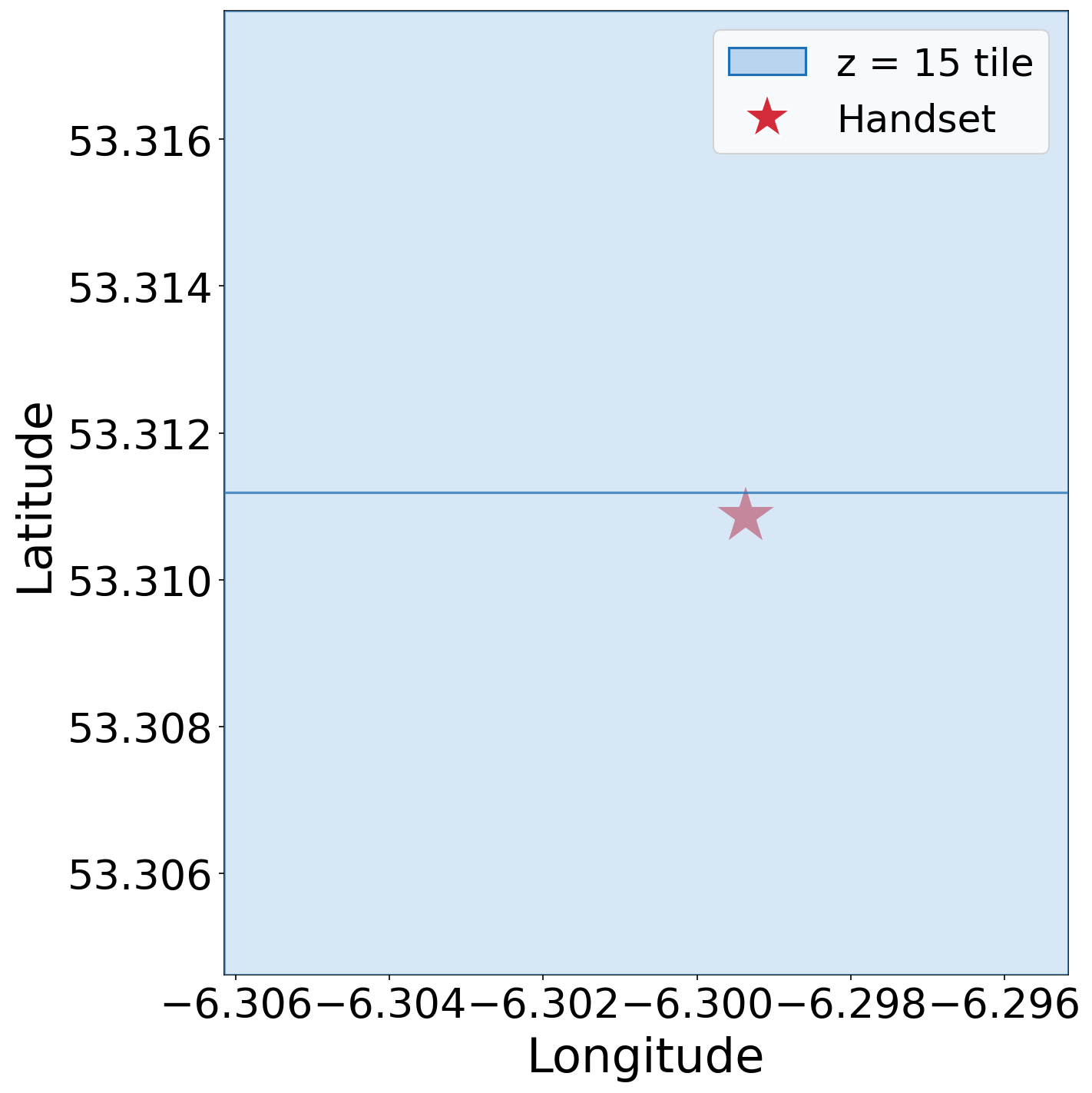}
\subcaption{Tile styles 1 and 11}
\end{subfigure}
\begin{subfigure}[b]{0.45\columnwidth}
\includegraphics[width=\columnwidth]{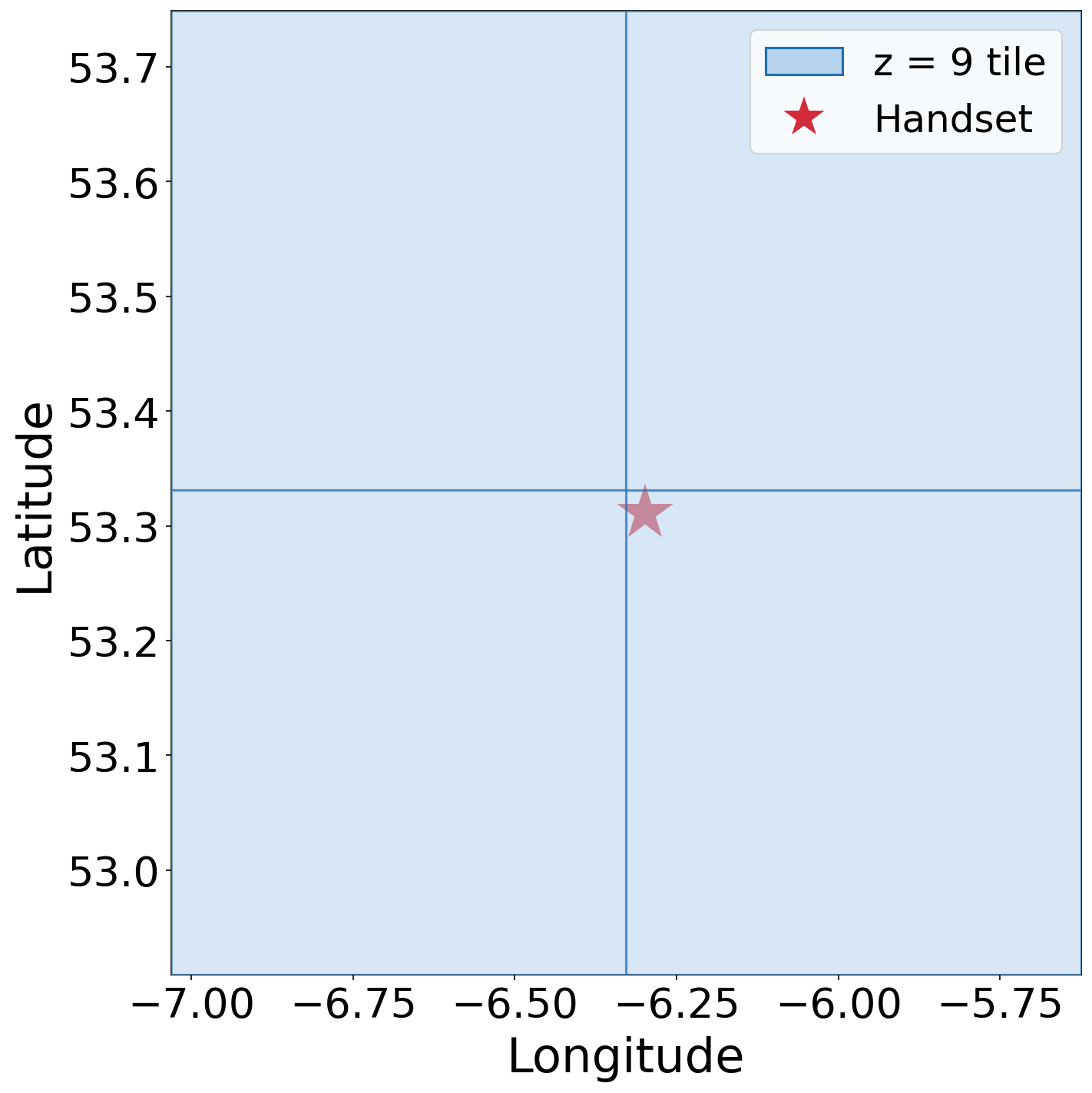}
\subcaption{Tile styles 20 and 68}
\end{subfigure}

\vspace{1mm}
\begin{subfigure}[b]{0.45\columnwidth}
\includegraphics[width=\columnwidth]{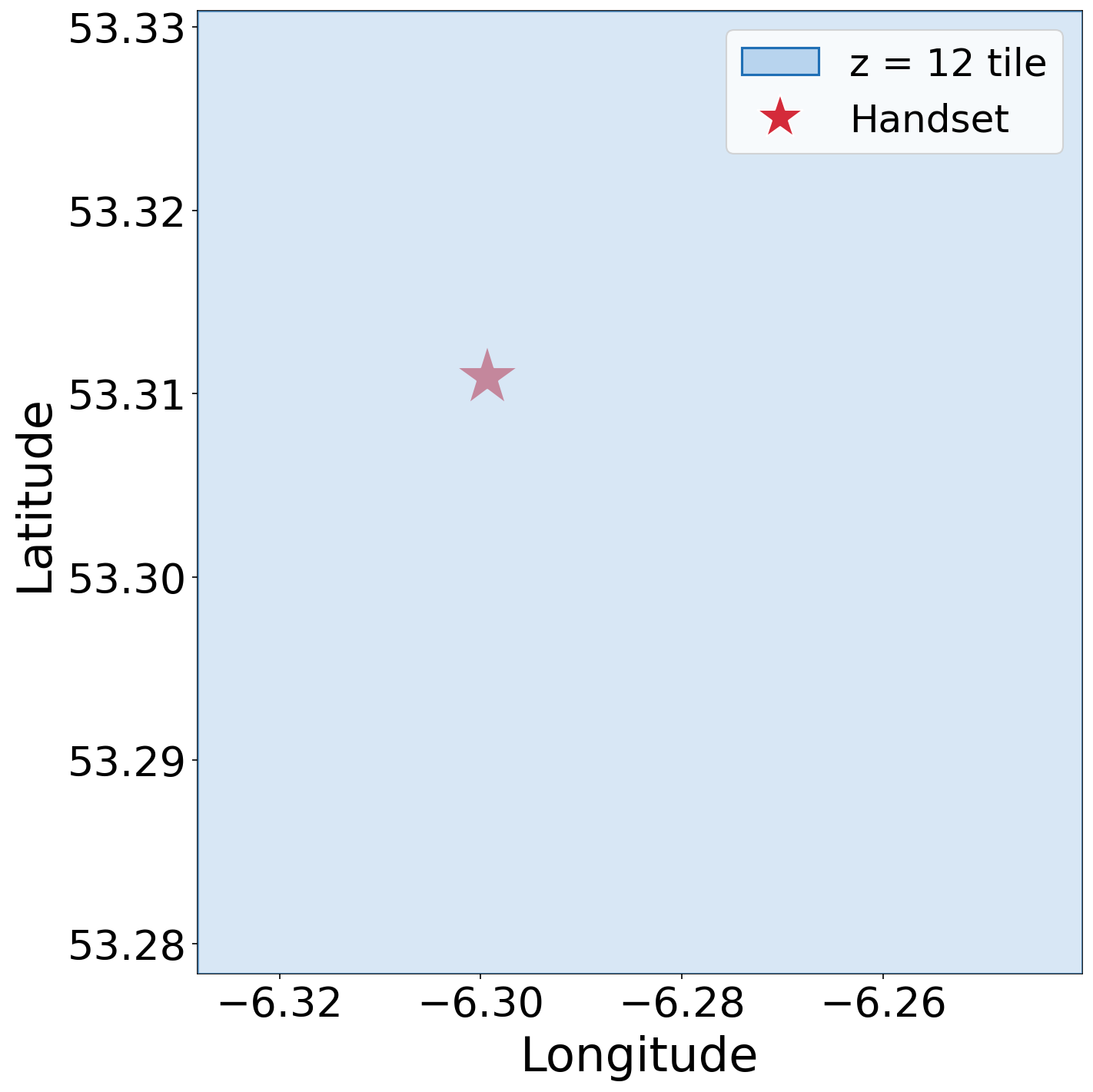}
\subcaption{Tile style 62}
\end{subfigure}
\begin{subfigure}[b]{0.45\columnwidth}
\includegraphics[width=\columnwidth]{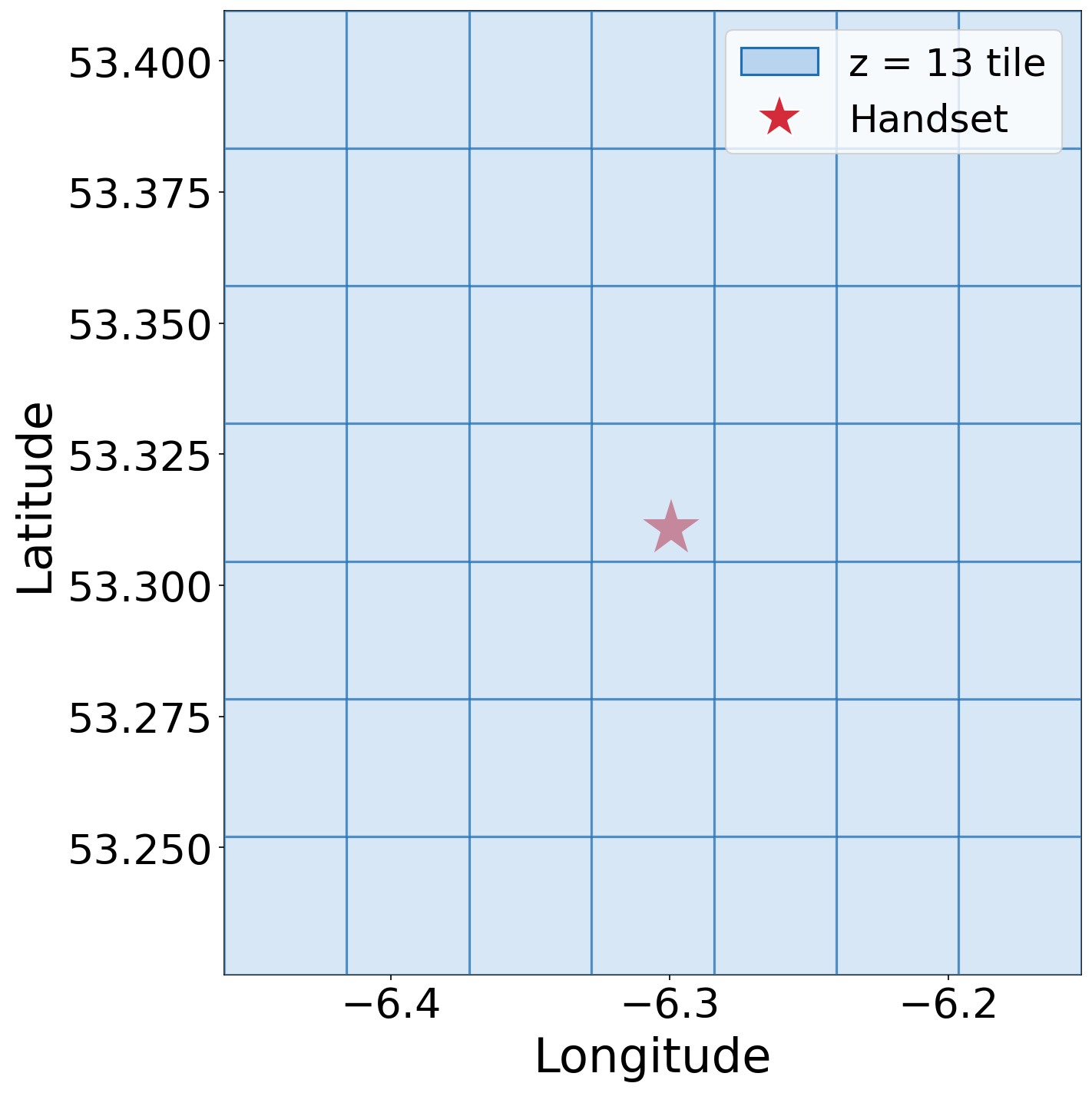}
\subcaption{Wi-Fi tiles}
\end{subfigure}

\caption{Visualisation of highest zoom-level tiles of each style downloaded by geod (the style is specified in the download request (inspection of the GeoServices code maps the style number to enum values e.g. style=1 is \nolinkurl{VECTOR_STANDARD}, style=11 is \nolinkurl{VECTOR_BUILDINGS}).  The star in each plot marks the handset location.}\label{fig:geod_tiles}
\end{figure}

Figure \ref{fig:geod_tiles} and, in the Additional Material, Figure \ref{fig:locd_tiles} show the smallest tiles that each process downloads, with the handset location marked by a star. When a single tile is downloaded it contains the handset (Figure \ref{fig:geod_tiles}(c)), so its location is revealed to an accuracy of about the tile width. When several tiles are downloaded they are centred on the handset (Figures \ref{fig:geod_tiles}(a),(b),(d) and \ref{fig:locd_tiles}(a),(b)), so averaging their positions locates it more accurately than the tile width\footnote{In Figure \ref{fig:locd_tiles}(c) the handset is in the left of the two tiles downloaded, presumably because our measurements ended before the tile to its left was downloaded.}.

With location off, \texttt{geod} still downloads map tiles but only at zoom levels 2 and 3, each covering thousands of km, which reveals no more than the location that can be inferred from the handset IP address.

\subsubsection{\texorpdfstring{Location Sent For \texttt{geod} Clients}{Location Sent For geod Clients}}\label{location-sent-for-geod-clients}

\texttt{geod} makes requests for several clients, and each request names its client. Table \ref{tab:geod-clients} shows the location carried by each client's requests, and by the direct requests of \texttt{photoanalysisd} and \texttt{mapspushd}. Apart from CoreRoutine, the Wi-Fi service and \texttt{photoanalysisd}, the location sent is coarse or absent. The Wi-Fi service sends no coordinates, but the MAC address of the connected access point locates the handset to tens of metres if Apple holds the position of the access point.

\begin{table}[htb]
\centering
\scriptsize
\setlength{\tabcolsep}{3pt}
\caption{Location sent to Apple's geo servers, by client.  The first seven clients send through \texttt{geod}, which names the client in each request.  The last two, marked direct, connect to the servers themselves.}\label{tab:geod-clients}
\begin{tabular}{|p{2.5cm}|p{2.2cm}|p{3.4cm}|}
\hline
\textbf{Client} & \textbf{Request} & \textbf{Location sent} \\
\hline
Weather widget & Reverse geocoding & City-level (one fixed point several kilometres from the handset; location-off experiment only) \\
\hline
\texttt{analyticsd} & Reverse geocoding & Coordinates rounded to 1 to 2 decimal places (about 1 to 10 km) \\
\hline
\texttt{network-}\newline\texttt{serviceproxy} & Reverse geocoding & Reported accuracy about 3.7 km \\
\hline
\texttt{kbd} & Geocoding & Postal district, no coordinates \\
\hline
CoreLocation prompt plugins & Dataset status check & None (IP address only) \\
\hline
Wi-Fi service (\texttt{ThreeBars-}\newline\texttt{XPCService}) & Wi-Fi quality lookup & MAC address of the connected access point, no coordinates \\
\hline
CoreRoutine & Reverse geocoding; Wi-Fi fingerprint lookup & Precise (reported accuracy 5.6 to 30 m in the lookups that state one); also nearby access points and visit times \\
\hline
\texttt{photoanalysisd}\newline(direct) & Batch reverse geocoding; spatial lookup & Precise (7 decimal places; handset position in our captures); search radius 1 to 20 km \\
\hline
\texttt{mapspushd}\newline(direct) & Announcements fetch & None \\
\hline
\end{tabular}
\end{table}

\subsubsection{Locations Identified By Wi-Fi Or Cell Identifiers}\label{locations-identified-by-wi-fi-or-cell-identifiers}

\texttt{locationd} sends a Wi-Fi MAC address or a cell tower identifier to \texttt{gs-loc.apple.com/clls/wloc} (\texttt{wloc} for short), and the response lists nearby access points or towers and their positions (Figure \ref{fig:wloc_wifi}; Figure \ref{fig:wloc_cell} in the Additional Material shows a cell example). Provided that Apple knows the position of the access point or tower, the identifier locates the handset to tens of metres (Wi-Fi) or hundreds of metres (cell). We observe 37 \texttt{wloc} requests, all with location on. One request instead carries the handset's precise latitude and longitude and the response lists 400 nearby access points. The Wi-Fi quality lookup (\nolinkurl{gsp85-ssl.ls.apple.com/wifi_request}), sent by \texttt{geod} for the Wi-Fi prefetch service \texttt{com.apple.wifi.ThreeBarsXPCService}, sends the MAC address of the access point that the handset is using, and we observe this even with location off.

\begin{figure}[htb]
\centering
\begin{lstlisting}[frame=single,language=http]
POST https://gs-loc.apple.com/clls/wloc
!{\color{blue}Request Headers:}!
User-Agent : locationd/2787.0.5 CFNetwork/1410.1 Darwin/22.6.0
!{\color{blue}POST Body (decoded):}!
apple arpc header: version=1, locale=en_US, app=com.apple.locationd, osVersion=16.7.11.20H360, requestType=1, length=71
wirelessAPs {
  macID: "dc:a6:32:fc:98:64"
}
numberOfSurroundingWifis: 100
surroundingWifiBands: K2DOT4_GHZ
wifiAltitudeScale: K_WIFI_ALTITUDE_SCALE10TO_THE2
meta {softwareBuild: "iPhone OS16.7.11/20H360", productId: "iPhone10,1"}
!{\color{blue}Response Data:}!
wirelessAPs {
  macID: "dc:a6:32:fc:98:64"
  location {
    latitude: 53.31<redacted>, longitude: -6.29<redacted>
    accuracy: 20, locationType: 3, altitude: 4812
    verticalAccuracy: 1000, score: 53, reach: 89
  }
  channel: 11
}
<more Wi-Fi MAC addresses, their lat/lon locations and coverage>
\end{lstlisting}
\caption{Example showing transmission of a Wi-Fi MAC address to Apple server \nolinkurl{gs-loc.apple.com/clls/wloc}.}\label{fig:wloc_wifi}
\end{figure}

\subsubsection{Precise Location}\label{precise-location}

We call a location precise when its coordinates have not been coarsened, which is Apple's term for the opposite of an approximate location. Precise locations are sent in all five sensor data uploads (Section \ref{sec:harvest}), in the \texttt{wloc} request that carries the handset position and in two kinds of lookup request that \texttt{geod} makes for CoreRoutine (all observed only with location on):

\begin{itemize}
\tightlist
\item
  Reverse geocoding requests to \nolinkurl{gsp57-ssl-revgeo.ls.apple.com/dispatcher.arpc}, sent by \texttt{geod}, carry a precise location, and the response includes the street address (Figure \ref{fig:dispatcher}). The request names CoreRoutine as the client.
\item
  Wi-Fi fingerprint lookups to \nolinkurl{gsp57-ssl-locus.ls.apple.com/dispatcher.arpc}, sent by \texttt{geod} for CoreRoutine, carry 94 locations, the MAC addresses and signal strengths of nearby access points and the entry and exit times of a visit. The purpose of the request is not clear.
\end{itemize}

\begin{figure}[htb]
\centering
\begin{lstlisting}[frame=single,language=http]
POST https://gsp57-ssl-revgeo.ls.apple.com/dispatcher.arpc
!{\color{blue}Request Headers:}!
User-Agent : geod/1 CFNetwork/1410.1 Darwin/22.6.0
!{\color{blue}POST Body (decoded):}!
apple arpc header: version=1, locale=en-IE_IE, app=com.apple.geod, osVersion=16.7.11.20H360, requestType=60, length=832
analyticMetadata {
  appIdentifier: "com.apple.CoreRoutine.helperservice"
  <...>
  sessionId {low: 7127060154989499779, high: 11333686820298006960}
  serviceTags {serviceType: GEO_SERVICES_FRAMEWORK, tag: "90E72145-<redacted>"}
}
clientMetadata {
  deviceCountryCode: "IE", deviceDisplayLanguage: "en-IE"
  <...>
  clientRevisions: AZUL_RELEASE
  <...>
  abClientMetadata {clientDatasetMetadata {bucketId: 7428}}
}
displayLanguages: "en-IE"
requestedComponents {type: ENTITY, count: 1}
<...>
requestType: REVERSE_GEOCODING
placeRequestParameters {reverseGeocodingParameters {extendedLocations {latLng {
  lat: 53.31<redacted>, lng: -6.29<redacted>
}}}}
!{\color{blue}Response Data:}!
<reverse geocoding data, including house address>
\end{lstlisting}
\caption{Example showing transmission of a precise lat/lon location to Apple server \nolinkurl{gsp57-ssl-revgeo.ls.apple.com/dispatcher.arpc}.}\label{fig:dispatcher}
\end{figure}

\subsubsection{Location Off}\label{location-off}

During the 27 hours of the experiment with location off, \texttt{locationd} made 7 connections (GNSS assistance data and Bluetooth configuration) and sent no location. \texttt{geod} made 299 connections, including configuration and resource downloads, telemetry, A/B experiment assignment, map tile downloads at zoom 2 and 3, one weather widget lookup with a city-level location and 97 Wi-Fi quality lookups. Each Wi-Fi quality lookup sends the MAC address of the access point that the handset is connected to. \texttt{wloc} lookups, \texttt{locationd} tile downloads, higher zoom \texttt{geod} tile downloads and all uploads occur only with location on. Switching location off therefore limits the location sent to city level, apart from the MAC address of the connected access point.

\subsection{Apple's ``Harvesting'' Of Phone Sensor Data}\label{sec:harvest}

\texttt{locationd} collects data from the handset sensors and uploads it to Apple servers, either itself or, for the real-time traffic probe, through \texttt{geod}. The \texttt{locationd} code refers to this collection as ``harvesting''. In brief:

\begin{enumerate}
\def\labelenumi{\arabic{enumi}.}
\tightlist
\item
  \emph{A wide range of sensor data is collected.} This includes cell and Wi-Fi observations, GPS locations with satellite measurements, pressure, temperature, compass attitude, step counts and motion state.
\item
  \emph{The user is not asked about it.} No consent is requested during setup, and declining to share iPhone Analytics does not stop it.
\item
  \emph{We observe it only in the drive experiment, and not always with an app.} No uploads occur with location off, nor in the static experiment with location on. The GPS uploads (\texttt{/pds/pd}, \texttt{rtloc}) cover only the minutes when the Apple Maps app was open and name it as the app using location. One trace upload (\texttt{/hvr/trc}), however, contains GPS fixes that name no app and inspection of the \texttt{locationd} code suggests that it can start GNSS for harvesting without an app requesting location\footnote{The monitor that receives the GPS fixes for the harvesting code is \texttt{CLMonitorGps}, which subscribes to the GPS location updates inside \texttt{locationd}; it applies no app or client condition to the fixes. \texttt{CLPolicyProactive} can start GNSS without a client app, but only when several conditions hold, among them that the handset is plugged in and charging, that the battery is at least 30\% (the default threshold) and that the Traffic system service is authorised.}.
\item
  \emph{Data is held and sent in batches.} Uploads are up to 13 hours old when sent, apart from the real-time traffic probe sent by \texttt{geod}, which is sent within a minute.
\end{enumerate}

Table \ref{tab:harvest} lists the uploads that we observe. The data is paired with the handset location, which has accuracy from 5 m (GPS) to 400 m (cell and Wi-Fi) for most records and up to 2.6 km for some in the trace uploads. Uploads include the handset hardware model and OS build and, in some, the identifier of the app using location and a \texttt{tripId} value that links them (Section \ref{sec:identifiers}), but we did not observe any user or device identifiers.

\begin{table*}[htb]
\centering
\scriptsize
\setlength{\tabcolsep}{3pt}
\caption{Sensor data uploads observed.  All but \texttt{rtloc} are sent by \texttt{locationd}.}\label{tab:harvest}
\begin{tabular}{|p{4.6cm}|p{8.6cm}|p{4.2cm}|}
\hline
\textbf{Endpoint} & \textbf{Data} & \textbf{Uploads observed} \\
\hline
\nolinkurl{gsp10-ssl.apple.com/hcy/pbcwloc} & LTE cell (MCC, MNC, TAC, cell ID, band, signal strength) and Wi-Fi (MAC address, channel, signal strength, scan time) observations, each with the handset location and motion state & 5 in two batches 12 hours apart, data up to 12 hours old \\
\hline
\nolinkurl{gsp10-ssl.apple.com/pds/pd} & GPS locations with altitude, accuracy, speed, course and motion state, app identifier, \texttt{tripId} & 1, with 99 locations 13 hours old \\
\hline
\nolinkurl{gsp9-ssl.apple.com/hvr/v2/rtloc} (sent by \texttt{geod}) & As \texttt{/pds/pd} & 1, with 66 locations 42 s old \\
\hline
\nolinkurl{gsp10-ssl.apple.com/hvr/trc} & Pressure, temperature, locations with GNSS satellite reports (satellite identifier, signal strength, azimuth, elevation), compass attitude, pedometer, motion state, app identifier & 2, with 317 and 864 events, up to 13 hours old \\
\hline
\nolinkurl{gsp10-ssl.apple.com/psr} & Barometric pressure with the GPS fix at the time of the reading & 3, one with an empty body \\
\hline
\end{tabular}
\end{table*}

The \texttt{rtloc} upload, sent by \texttt{geod}, carries 66 locations at 1 s intervals with accuracy about 5 m and the motion state (Figure \ref{fig:rtloc} in the Additional Material). The locations are those that \texttt{locationd} also uploads in \texttt{/pds/pd}, which carries 99 locations over 7.5 minutes with the bundle identifier of the app using location (Figure \ref{fig:pdspd}). Code inspection indicates that \texttt{locationd} passes the locations to GeoServices, which hands them to \texttt{geod} for upload. The \texttt{tripId} value is the same in both uploads, so they can be linked (Section \ref{sec:identifiers}).

Collection and transmission are decoupled. The \texttt{rtloc} upload follows the last location by 42 s. The other uploads are later: observations in \texttt{pbcwloc} are up to 12 hours old, \texttt{/pds/pd} and \texttt{/hvr/trc} 13 hours\footnote{In \texttt{locationd}, the trace code stores \texttt{fmod(traceStartTime,\ interval)} and sets each event's \texttt{timestamp} to its time minus this offset. The setting \texttt{HarvestIndoorTraceFuzzInterval} (alternatively \texttt{traceFuzzInterval}) defaults to 900 s and a configuration dictionary can override it. The data indicate an interval of one day: matching \texttt{/pds/pd} fixes show the true start was 12.7 hours before upload, and the recorded event times are multiples of 86,400 s from UTC midnight. The age given allows for this. For one trace the true times follow from matching its GPS fixes to those in \texttt{/pds/pd}, and the other is at most 5 hours old.}, and the CoreRoutine Wi-Fi fingerprint lookups are about 2 days after the visit. Apple therefore receives a history of where the handset has been, not only its current position. We have not established what triggers each upload.

\begin{figure}[htb]
\centering
\begin{lstlisting}[frame=single,language=http]
POST https://gsp10-ssl.apple.com/pds/pd
!{\color{blue}Request Headers:}!
User-Agent : locationd/2787.0.5 CFNetwork/1410.1 Darwin/22.6.0
!{\color{blue}POST Body (decoded):}!
apple arpc header: version=1, locale=en_US, app=com.apple.locationd, osVersion=16.7.11.20H360, requestType=101
meta {productId: "D20AP", softwareVersion: "iPhone OS16.7.11/20H360"}
locations {
  latitude: 53.37<redacted>
  longitude: -6.59<redacted>
  horizontalAccuracy: 5, altitude: 61.95, verticalAccuracy: 4
  timestamp: 802283024.0005983
  provider: 1, appBundleIdIndices: "0"
  rawMotionActivity {type: IN_VEHICLE_STATIC, confidence: HIGH}
  motionActivity {type: IN_VEHICLE_STATIC, confidence: HIGH}
  dominantMotionActivity {type: IN_VEHICLE_STATIC, confidence: HIGH}
}
<similar entries with handset location at different times>
locations {
  latitude: 53.37<redacted>
  longitude: -6.59<redacted>
  horizontalAccuracy: 5, altitude: 63.45, verticalAccuracy: 3
  speed: 4.18, course: 110.4, courseAccuracy: 6.66, speedAccuracy: 0.84
  timestamp: 802283043.0001837
  provider: 1, appBundleIdIndices: "0"
  rawMotionActivity {type: DRIVING, confidence: HIGH}
  motionActivity {type: DRIVING, confidence: HIGH}
  dominantMotionActivity {type: IN_VEHICLE_STATIC, confidence: HIGH}
}
<similar entries with handset location at different times>
appBundleId: "com.apple.Maps"
tripId: "7B0447BD-<redacted>"
appBundleIds: "com.apple.Maps"
\end{lstlisting}
\caption{Example showing transmission of GPS fixes, with the handset motion state, to Apple server \nolinkurl{gsp10-ssl.apple.com/pds/pd}.  The two fixes shown are also sent to \nolinkurl{gsp9-ssl.apple.com/hvr/v2/rtloc} (Figure \ref{fig:rtloc}).}\label{fig:pdspd}
\end{figure}

Presumably Apple uses the data to build and refresh its Wi-Fi, cell, traffic and barometric databases, but we have not established this. The user is not asked about this collection during setup (Table \ref{tab:settings} in the Additional Material), and the iPhone Analytics option, which the test user set to ``Don't Share'', does not prevent it. We did not test whether switching off a setting under Location Services (for example Routing and Traffic) stops it, so we cannot say whether any opt out exists short of switching location off, which our measurements suggest stops it.

\subsection{Telemetry \& Experiments}\label{sec:telemetry}

\texttt{geod} sends usage records to Apple and receives an A/B experiment assignment from it. In brief:

\begin{enumerate}
\def\labelenumi{\arabic{enumi}.}
\tightlist
\item
  \emph{Telemetry is sent with location off and with iPhone Analytics declined.} The test user chose ``Don't Share'' at setup.
\item
  \emph{The records describe each request that \texttt{geod} makes and the client that made it.}
\item
  \emph{A daily record carries an identifier that persists for months.}
\item
  \emph{The handset is assigned to A/B experiments, and the bucket identifier is echoed to Apple in later requests.}
\end{enumerate}

\subsubsection{Telemetry}\label{telemetry}

\texttt{geod} sends batches of log records to \texttt{gsp64-ssl.ls.apple.com/hvr/v3/use}. The request is a protobuf and the response is an empty acknowledgement. We observe 88 connections in the 27 hours with location off, 52 in the 16 days with location on and the handset stationary, and 37 in the drive experiment.

Most records describe a network request made by \texttt{geod}. Each gives the network service used (for example manifest, resource download, A/B assignment, Wi-Fi quality or place data), the application that requested it, its duration, the request and response sizes, the HTTP status or error code, the server IP address and port and the tile set identifier. The requesting application identifier can include the calling class and method. For example, a CoreRoutine request to identify the place of a visit from a Wi-Fi fingerprint is recorded with the method name. Records also carry the handset model, OS version, country code, carrier name, connectivity type (Wi-Fi or cellular) and whether cellular data is allowed. The network event records contain no location.

Once a day \texttt{geod} also sends a daily use summary and a device sync status record, which the protobuf labels \texttt{FIFTEEN\_MONTH\_SESSION} and \texttt{FIFTEEN\_MONTH\_USER\_SESSION}. Both carry session IDs (Section \ref{sec:identifiers}), and the labels suggest that Apple retains the records for 15 months. We did not observe daily records in the 27 hours with location off.

The other telemetry records also carry a session ID (see Section \ref{sec:identifiers}).

\subsubsection{A/B Experiments}\label{ab-experiments}

\texttt{geod} sends a request to \texttt{gsp-ssl.ls.apple.com/ab.arpc} containing a random \texttt{guid} value, the locale, OS version and country code, and receives an A/B experiment bucket identifier for each of four experiment namespaces (Parsec, Siri, RAP and Maps). The bucket identifier is the same for all four. The response contains a refresh interval of 4 hours (most of the requests that we observe are 4 hours apart) and a branch expiration time of 30 days, presumably the lifetime of the assignment. The handset echoes the bucket identifier to Apple in \texttt{dispatcher.arpc} requests, presumably so that Apple can attribute behaviour to an experiment group.

\subsection{Analysis Of Identifiers Used}\label{sec:identifiers}

We distinguish three kinds of identifier. A ``direct identifier'' identifies the individual person or handset, e.g.~an Apple Account ID or a device serial number. A ``session identifier'' is a random value created by the handset's software and reused across records for a period of time, so that records can be associated with one another. Examples below include the telemetry session IDs, the A/B \texttt{guid} and the \texttt{tripId}. A ``network identifier'' comes from the network the handset is using and not from its software, e.g.~the IP address or the MAC address of a nearby Wi-Fi access point. It potentially links connections for as long as the handset stays on that network.

\subsubsection{Direct Identifiers}\label{direct-identifiers}

A direct identifier sent alongside data often makes that data ``personal data''\footnote{GDPR Article 4(1) says personal data is ``any information relating to an identified or identifiable natural person''.}. We did not observe any direct identifiers in \texttt{geod} or \texttt{locationd} connections, including when the user is logged in to an Apple account.

\subsubsection{Session Identifiers}\label{session-identifiers}

A session identifier is more complicated since its impact depends, amongst other things, on: (i) the data that it links together, and (ii) the lifetime of the identifier (which affects the duration over which data is linked together). For example, linking location data together over several days might easily identify an individual, e.g.~from where they spend the night (which shows where they live) and where they spend the day (which shows where they work). A short-lived session identifier, e.g.~one that persists for a few seconds or minutes, is generally much less of a concern than a long-lived identifier which persists for days or months. Additionally, we might be concerned about whether a session identifier can be easily linked to a direct identifier, e.g.~if the two are sent in the same network connection. However, no direct identifiers are sent in \texttt{geod} or \texttt{locationd} connections and so this is not a concern here.

\begin{table}[tb]
\centering
\scriptsize
\setlength{\tabcolsep}{3pt}
\caption{Identifiers observed in \texttt{geod} and \texttt{locationd} connections and whether Apple's documentation describes them.  Lifetimes are measured from the times at which records are sent, and from the session age field in the telemetry.}\label{tab:identifier-docs}
\begin{tabular}{|p{2.2cm}|p{3.3cm}|p{2.0cm}|}
\hline
\textbf{Identifier} & \textbf{Lifetime observed} & \textbf{Documented} \\
\hline
Daily record session IDs & At least 106 days & No; the 15-month label is not mentioned \\
\hline
Session ID in \texttt{geod} telemetry & 20 IDs across 11 processes.  Maximum age per ID 3 to 29 days.  Used over 16 to 17 days by \texttt{geod}, the Wi-Fi quality service and a prompt plugin & No \\
\hline
\texttt{dispatcher.arpc} \texttt{sessionId} & 10 IDs across the clients, none seen for more than 10 seconds & No \\
\hline
A/B \texttt{guid} & At least 1.8 days; a different value in the drive experiment & No \\
\hline
A/B \texttt{bucketId} & Assigned by server: 7428 from 17 February to 6 March, 12296 on 4 and 5 June & No \\
\hline
\texttt{tripId} & One trip & Yes, rotates when the trip ends\\
\hline
MAC address of connected access point & Same address in all 184 Wi-Fi quality lookups (17 February to 6 March); also seen in three other endpoints up to 5 June, 108 days & No \\
\hline
MAC addresses of neighbouring access points & Sent in \texttt{wloc} lookups, \texttt{pbcwloc} uploads and the CoreRoutine Wi-Fi fingerprint lookups; six access points occur in both an upload and a lookup sent up to 12 hours apart & No \\
\hline
IP address & Not recorded (Apple sees it in every connection) & Yes, for approximate location \\
\hline
\end{tabular}
\end{table}

Table \ref{tab:identifier-docs} lists the identifiers that we observe in \texttt{geod} or \texttt{locationd} connections.

\emph{Session IDs in the daily records.} The daily device sync status record lists two session IDs, the \texttt{sessionId} and \texttt{preSyncSessionId}, both created on 19 February at 04:31 UTC (\texttt{createTime}). Both are still in use in the records sent on 4 and 5 June, 106 days later. The \texttt{FIFTEEN\_MONTH\_SESSION} summary carries a third ID that is unchanged from 19 February to 5 June. We do not know when these values change, but the record labels suggest that Apple retains the records for 15 months.

\emph{Session IDs in telemetry.} Each record in \texttt{geod}'s usage telemetry (Section \ref{sec:telemetry}) carries a session ID and its age (\texttt{ageOfSessionIdInSeconds}). The IDs come from 11 processes\footnote{\texttt{geod} itself, the Wi-Fi quality service, CoreRoutine, \texttt{locationd}, \texttt{analyticsd}, \texttt{kbd}, the Weather widget, \texttt{photoanalysisd}, \texttt{networkserviceproxy} and the two permission-prompt plugins.} and several persist for over two weeks: those of \texttt{geod}, the Wi-Fi quality service and one permission-prompt plugin are each used in records sent over 16 to 17 days.

\emph{Session IDs in \texttt{dispatcher.arpc} requests.} Each client sends a session ID in its \texttt{dispatcher.arpc} requests. For the system processes it differs from their telemetry session ID. We observe 10 such IDs. Each is shared by requests made close together (the three Wi-Fi fingerprint lookups share one) and none is seen for more than 10 seconds, but we have too few requests to establish how long they last.

\emph{A/B experiment identifiers.} The \texttt{guid} value sent to \texttt{gsp-ssl.ls.apple.com/ab.arpc} is random. We observe one value in all 12 requests of the first two parts of the experiment (17 to 19 February) and a different value in the 7 requests of the drive experiment (4 to 5 June), so it can link requests to that endpoint over at least 1.8 days. The A/B experiment \texttt{bucketId} (Section \ref{sec:telemetry}) is assigned by the server and so is probably shared by many handsets, but we cannot establish this from measurements on a single handset. The service tag in \texttt{dispatcher.arpc} requests is a random UUID that differs for each connection.

\emph{Trip identifier.} The \texttt{tripId} value is a random UUID sent in both the \texttt{rtloc} upload by \texttt{geod} and the \texttt{/pds/pd} upload by \texttt{locationd} made 13 hours later, so it links the transmissions of the two processes even when there is a delay between uploads. We observe a single trip. Inspection of the \texttt{locationd} code indicates that it creates a new value when no location has been collected for 15 minutes, or when a trip lasts more than 3 hours\footnote{\texttt{CLPolicyTrip::createNewTripId()} generates the UUID, and the function that decides when to call it creates a new value if none exists, if more than 900 s have passed since the last location accepted by either the Tracks or the Realtime harvester, or if the trip has lasted more than \texttt{MaxTripDuration}, which defaults to 3 h.}.

While the telemetry records contain no location, they can potentially be matched to the requests that do. A record for a CoreRoutine request gives the request's size in bytes, its type and the method that made it. For each of the nine CoreRoutine requests in our captures, exactly one record has a size equal to the request's length and was sent within 15 minutes after it (5,393, 3,372 and 997 bytes for the three Wi-Fi fingerprint lookups). The match needs no shared identifier, and Apple receives both from the same IP address. The telemetry's session ID attaches to a request that carries a precise location, and since the telemetry session IDs persist for several days, they could link requests made days apart. We do not know whether Apple makes this association, and indeed it seems unlikely that it does, but we note that the possibility exists. This would be mitigated by Apple using shorter-lived telemetry session IDs.

\subsubsection{Network Identifiers}\label{network-identifiers}

The IP address is necessarily revealed in every network connection, and potentially links all connections made from the same network\footnote{Multiple devices may share the same network, and so the same IP address, in which case the linkage is for that group of devices.}. The MAC address of the connected access point is sent in every Wi-Fi quality lookup, with the same address in all of the lookups in the location-off experiment, and it appears in four endpoints over 108 days (190 requests). The MAC addresses of neighbouring access points are sent in \texttt{wloc} lookups (164 distinct addresses over the 29 requests that carry a MAC address), in the \texttt{pbcwloc} uploads and in CoreRoutine's Wi-Fi fingerprint lookups. The same access point therefore links these records: six appear in both the \texttt{pbcwloc} uploads and the Wi-Fi fingerprint lookups, which were sent up to 12 hours apart.

\subsubsection{Values Shared By Many Handsets}\label{values-shared-by-many-handsets}

Requests carry the handset model (\texttt{iPhone10,1}, or \texttt{D20AP} in uploads), OS build, locale and country code. The \texttt{accessKey} URL parameter and the \texttt{maps-auth-token} header sent in tile and resource downloads are not individual device identifiers (see Section \ref{sec:access-key} in the Additional Material).

\subsection{Comparison With Apple's Public Documentation}\label{sec:docs}

Apple's public documentation of Location Services does not describe network connections, endpoints or message contents, so it cannot be compared with our measurements connection by connection. It does make statements about what data is sent to Apple and about the identifiers attached to it, and these can be checked.

We compare our observations with the Location Services Privacy Overview white paper of November 2019 \citep{apple_location_whitepaper}, and with the Apple Privacy Policy and the ``Location Services \& Privacy'', ``Apple Maps \& Privacy'' and ``Device Analytics \& Privacy'' pages\footnote{\texttt{apple.com/legal/privacy/en-ww} and \texttt{apple.com/legal/privacy/data/en/}\allowbreak\texttt{\{location-services,}\allowbreak\texttt{apple-maps,}\allowbreak\texttt{device-analytics\}}. Apple does not date these pages, so we use Internet Archive snapshots taken while the experiments were running (31 January, 11, 13 and 17 February, and 30 and 31 May 2026). The quoted text is the same in each snapshot of a page.} These pages are not versioned and apply to every supported device. We cannot check statements about server-side handling, such as aggregation.

For each statement we give its context, what we observe, and an assessment: \emph{consistent}; \emph{incomplete} (the statement is not contradicted but omits what we observe); \emph{not addressed} (the documentation does not cover what we observe); or \emph{contradicted}. Omissions are not contradictions, and we report a contradiction only where a statement is incompatible with what we observe. Table \ref{tab:docs-summary} summarises the assessments and Table \ref{tab:identifier-docs} lists the identifiers that we observe.

\begin{table}[tb]
\centering
\scriptsize
\setlength{\tabcolsep}{3pt}
\caption{Summary of the comparison of Apple's statements with our observations.}\label{tab:docs-summary}
\begin{tabular}{|p{5.6cm}|p{2.6cm}|}
\hline
\textbf{Apple statement} & \textbf{Assessment} \\
\hline
1. Wi-Fi and cell submissions are anonymous, encrypted and use a rotating identifier & Incomplete \\
\hline
2. Routing and Traffic data is sent with an identifier that rotates at the end of a trip & Consistent for the identifier; incomplete for the data \\
\hline
3. With Location Services off, apps cannot access location & Incomplete \\
\hline
4. Device Analytics sharing can be disabled altogether & Not addressed \\
\hline
5. Significant Locations are encrypted end to end; on-device processing means Apple does not learn the pattern of life & Not addressed (encryption); contradicted (pattern of life) \\
\hline
6. Maps identifiers rotate or are longer-lived & Identifier lifetimes are given only for Maps: Not addressed \\
\hline
7. Precise location is collected only to support services, or where the user agrees & Not assessed \\
\hline
\end{tabular}
\end{table}

\textbf{1. Wi-Fi and cell location submissions.}
\emph{Apple says:} the data is sent ``in an anonymous and encrypted form'', and only when Location Services is on (Location Services page). The white paper describes this as crowdsourcing to improve Apple's database of Wi-Fi hotspot and cell tower locations, and says that ``the submission uses a random, rotating identifier so that Apple can't tie multiple submissions together''.
\emph{We observe:} five \texttt{pbcwloc} uploads, sent only with location on. We take ``encrypted'' to mean TLS, since every payload that we decode is a plain protobuf inside the TLS session, so Apple's servers can read it. The request bodies contain no identifier, so none rotates. Each record pairs the MAC address of an access point or the identity of a cell with the handset location (accuracy 400 m or better) and a timestamp. The uploads share a source IP address, and six of their access points recur in CoreRoutine's Wi-Fi fingerprint lookups sent up to 12 hours apart.
\emph{Assessment:} incomplete. The uploads carry no direct identifier, which seems an improvement on the rotating identifier that Apple describes. However, ``anonymous'' and ``Apple can't tie multiple submissions together'' seem too strong since the connections contain values that could link submissions.

\textbf{2. Routing and Traffic.}
\emph{Apple says:} while in transit the handset sends ``GPS data, travel speed and direction, and barometric pressure'', anonymously (Location Services page). The white paper adds that the identifier ``rotates at the conclusion of a trip'' and is not tied to an Apple ID.
\emph{We observe:} only the drive experiment produces these uploads. The \texttt{rtloc} and \texttt{/pds/pd} uploads carry a random \texttt{tripId}, and the \texttt{/hvr/trc} and \texttt{/psr} uploads carry no identifier of any kind. None carries an Apple Account identifier. The \texttt{tripId} rotates after a 15-minute gap or a 3-hour trip (Section \ref{sec:identifiers}), which is close to the ``conclusion of a trip''. The uploads also carry data that the description does not list: motion state, altitude, the bundle identifier of the app using location, GNSS satellite reports, compass attitude, pedometer data and temperature, and they are up to 13 hours old when sent.
\emph{Assessment:} consistent for the identifier; incomplete for the data.

\textbf{3. Location Services off.}
\emph{Apple says:} when Location Services is off, ``apps can't access the user's Location Services data at any time'' (white paper), and apps and websites ``may still use other information, such as the IP address'' (Location Services page). Both statements are about apps and websites, and neither mentions Apple's own daemons. The white paper separately notes that ``the networks users connect to'' ``might be used to track location'', in a passage on apps that scan for Wi-Fi networks and Bluetooth beacons to get around location controls.
\emph{We observe:} in the 27 hours with location off, \texttt{geod} makes 299 connections. They include 97 Wi-Fi quality lookups, each carrying the MAC address of the connected access point, which locates the handset to tens of metres if Apple holds the position of the access point, and a weather widget lookup with a city-level location.
\emph{Assessment:} incomplete. Apple's statements about apps are not contradicted, but they say nothing about what its own daemons send.

\textbf{4. Device Analytics.}
\emph{Apple says:} users can ``disable the sharing of Device Analytics altogether'' (Device Analytics page). The page does not say which Apple services this setting covers.
\emph{We observe:} the test user chose ``Don't Share'' at setup, and \texttt{geod} nevertheless sends 88 telemetry connections in the 27 hours with location off.
\emph{Assessment:} not addressed, because we cannot tell whether this telemetry is Device Analytics.

\textbf{5. Significant Locations.}
\emph{Apple says:} Significant Locations \& Routes ``are synced between your devices using end-to-end encryption and cannot be read by Apple'' (Location Services page). The white paper adds that processing them on the device avoids ``Apple or third parties learning sensitive details about your pattern of life''.
\emph{We observe:} CoreRoutine, which apparently implements Significant Locations, sends requests to Apple, through \texttt{geod}, that are not end-to-end encrypted, so Apple's servers can read them. Its reverse geocoding requests carry precise locations. Its Wi-Fi fingerprint lookups carry 94 timestamped locations, the MAC addresses of nearby access points and the entry and exit times of a visit. Apple therefore receives where the handset was and for how long.
\emph{Assessment:} the encryption statement concerns the synchronised store and says nothing about these requests, so it is not addressed. The claim that Apple does not learn the pattern of life is contradicted for the visits that we observe, assuming that CoreRoutine's requests are part of Significant Locations.

\textbf{6. Identifier lifetimes.}
\emph{Apple says:} the Apple Maps page, which concerns the Maps app, says that individual usage metrics are ``associated with an identifier that rotates multiple times per hour''. It adds that ``Longer-lived identifiers'' are used for ``application performance, aggregate usage metrics'' and to maintain the integrity of the service, and that these are not attached to an Apple Account. It gives no lifetime for the longer-lived identifiers. No other page that we examined gives identifier lifetimes.
\emph{We observe:} telemetry session IDs that persist for days to months, an A/B experiment \texttt{guid} that persists for at least 1.8 days, and the connected access point's MAC address, which appears in four endpoints over 108 days (Table \ref{tab:identifier-docs}). The session IDs come from 11 system processes and are sent without Maps being used, including in the location-off experiment, so they are not Maps data.
\emph{Assessment:} not addressed. The Maps page concerns Maps and so says nothing about these identifiers or their lifetimes.

\textbf{7. Apple Privacy Policy.}
We do not assess its statement that Apple collects ``precise location only to support services such as Find My or where you agree for region-specific services''. It is too open-ended to test: ``such as'' leaves the list of services open and ``where you agree'' is not defined. We observe precise locations sent only with Location Services on, which is consistent with a reading in which switching it on is the agreement, but the policy does not say so and the per-service settings that Apple documents suggest that agreement may be intended per service. We did not test those settings (Routing and Traffic, Improve Location Accuracy and Significant Locations), so we do not know whether they stop the uploads and requests that we observe, nor whether the telemetry and the Wi-Fi quality lookups have a control at all.

\section{AI Use}\label{ai-use}

Summary of main AI use:

\begin{itemize}
\item
  \emph{Section \ref{sec:AI}:} Claude Opus 4.8, Codex GPT Sol 5.6 used for recovery of protobuf descriptors from Apple binaries.
\item
  \emph{Section \ref{sec:extractor}:} Claude Opus 4.8/Sonnet 5 used to develop Extractor.
\item
  \emph{Section \ref{sec:network}:} Claude Sonnet 5 used to: (i) assist with ObjC binary code analysis to trace the lifecycle of the GUID, Maps-Auth-Token, accessKey, service\_tag, session\_id, clientRevisions, abClientMetadata, Gloria tile ID (X-tilekey header) values, i.e.~how they are calculated and when they change (validated using network measurements and by human code review); (ii) assist with code analysis of URL location encoding (e.g.~\texttt{cl2.apple.com/1/v1/\{lat0\}/\{lon0\}/\{lat\}\_\{lon\}.gz}) (validated by human code review and known handset location); (iii) assist with decoding headers of Wi-Fi, GSM fine cell, DEM (Digital Elevation Model), HRSE (High-Resolution Signal Environment) tiles to extract tile geographic boundaries (the download URLs also often specify this information, so this is primarily a cross-check).
\end{itemize}

\section{Limitations}\label{limitations}

The measurements reported here are inevitably limited in nature. In our experiments we only make fairly minimal use of the handset, although we do change the handset location and take it for a drive. We also assume a privacy-conscious user who answers ``no'' when asked to share data with Apple and who uses essentially no Apple location services\footnote{The Apple Maps app is briefly opened to confirm that the reported handset location is accurate.}. Behaviour that is infrequent, or that depends on usage we did not carry out, may not be observed. This means that the data collection that we report here is likely an underestimate: absence of evidence is not evidence of absence. We observe only what is sent, not how Apple stores or uses it.

In our tests we use a jailbroken handset. This is necessary to allow decryption of the HTTPS network traffic, as discussed in Section \ref{sec:mitm}. It is possible that the behaviour of a jailbroken handset differs from that of a non-jailbroken one. However, we have seen no evidence of this.

The decoding of the traffic depends on Extractor. Its validation tests use binaries compiled from an ObjC template that we reverse engineered, and so share our assumptions about Apple's code, and for Apple's own binaries the true schemas are unknown. We mitigate this by checking that decoded messages are consistent with the field names and with the handset's known position.

Probably the main limitation of our measurements is the use of an older version of iOS. This is unavoidable since there are no jailbreaks for later versions. Later versions may make different connections or send different data, and we cannot test this. However, we can compare message formats and this shows that the protobuf fields we observe in iPhone 8 (iOS 16.7.11) traffic are still declared, unchanged, in later releases. We ran Extractor on the GeoServices and CoreLocationProtobuf images of iOS 27 and macOS 27 and compared the recovered descriptors with those of iOS 16: all 739 geo and clp message fields seen in our captures (188 messages) have the same field number, name and type in both. Later releases add fields (177 in iOS 27, 173 in macOS 27, to these 188 messages) but remove none of the observed ones.

\section{Conclusion}\label{conclusion}

We measure the location data that an iPhone sends to Apple and, because the message formats are undocumented, evaluate how far AI can help to recover them.

Asking two frontier models to reconstruct Apple's protobuf schemas directly from its binaries gets most individual fields right but only three of the 17 schemas entirely right for each model. The errors cannot be detected without an independent reference: they leave the schemas internally consistent and fall mostly on rarely used fields, so testing against captured traffic does not reveal them. Having Claude write a tool whose output can be validated is, in contrast, reliable and much less work than recovering the schemas by hand. The condition is that ground truth exists that does not come from the model. Here we construct it by reverse engineering the template that Apple's protobuf classes follow and compiling schemas whose answers we know.

The decoded traffic shows that no message carries a direct identifier of the user or handset. With location on, however, the handset sends its precise position to Apple, in reverse geocoding and Wi-Fi fingerprint lookups that carry the entry and exit times of visits, in Wi-Fi and cell lookups, and in batched uploads of GPS fixes and of Wi-Fi and cell observations. Usage telemetry carries session identifiers that persist for at least 106 days and so can link a handset's records over months, and the MAC address of the connected access point, a network identifier, is sent when location is off. Apple's documentation does not mention the last two. Our measurements are of one handset used in a limited way, so our account of the data sent is likely an underestimate.

We recommend that Apple (i) document the telemetry session identifiers and limit their lifetime, (ii) state which data is sent when location is off, or stop sending it, (iii) make clear that Significant Locations processing involves requests to Apple's servers that carry precise locations and visit times, and (iv) ask users before ``harvesting'' phone sensor data.

\bibliographystyle{IEEEtran}
\bibliography{bibfile}

\section*{Additional Material}\label{additional-material}
\addcontentsline{toc}{section}{Additional Material}

\subsection{Comparison With Published Protobuf Descriptors}\label{comparison-with-published-protobuf-descriptors}

This section compares the descriptors that Extractor recovers with descriptors published by other researchers and written by hand, for Apple's location services \citep{acheong08} and for CloudKit and iMessage \citep{rustpush}. The field numbers agree wherever both define a field, and in every case where the published descriptor disagrees with Extractor on a field's type, meaning or structure, the iPhone's binaries agree with Extractor.

\subsubsection{Location Protobuf Descriptors}\label{sec:acheong-compare}

The repository accompanying \citep{acheong08} contains three protobuf descriptors written by hand from captured traffic. They cover the Wi-Fi and cell positioning request and response (\texttt{wloc}), the \texttt{pbcwloc} request, and the Wi-Fi tile response (\texttt{/wifi\_request\_tile}). We compare these with the descriptors that Extractor recovers from the iPhone 8 binaries. Messages are paired by endpoint and field number: their \texttt{AppleWLoc} (a single message used for both directions) with \texttt{ALSLocationRequest} and \texttt{ALSLocationResponse} in \texttt{locationd}, their \texttt{PbcWlocRequest} with \texttt{CLPCellWifiCollectionRequest} in \texttt{CoreLocationProtobuf}, and their \texttt{WifiTile} with \texttt{GEOWiFiQualitySearchResultTile} in \texttt{GeoServices}. Both sets of descriptors are then used to decode the same captured traffic: 66 \texttt{wloc} requests, 66 \texttt{wloc} responses, 34 \texttt{pbcwloc} requests and 226 Wi-Fi tiles.

Every captured body decodes under Extractor's descriptors with no unknown fields. The published descriptors leave unknown fields in every message type. They omit the \texttt{wloc} request fields that carry cell towers (1), the Wi-Fi band (31) and the altitude scale (32), the channel and Z-axis harvest count of each access point (21, 22), the app bundle identifier in each \texttt{pbcwloc} Wi-Fi entry (5), and several fields of the tile entries.

The field numbers agree wherever both descriptors define a field. Where they differ is in the declared types and in the field names. Since the protobuf wire format does not distinguish \texttt{int32} from \texttt{int64} or \texttt{uint32} for small values, and carries no names, these differences cannot be resolved from traffic alone. We therefore resolve each one against the iPhone 8 binaries. The serialisation code in \texttt{writeTo:} calls a writer routine for each field (for example \texttt{PBDataWriterWriteInt32Field} or \texttt{PBDataWriterWriteUint32Field}) and the Objective-C ivar encoding records the stored type (\texttt{i} for \texttt{int}, \texttt{I} for \texttt{unsigned\ int}, \texttt{Q} for \texttt{unsigned\ long\ long}, \texttt{B} for \texttt{bool}). Table \ref{tab:acheong-types} lists every declared-type difference. In every case the binary agrees with Extractor.

\begin{table*}[htb]
\centering
\footnotesize
\caption{Differences in declared field type between the published descriptors and Extractor, with the writer routine called in \texttt{writeTo:} and the ivar encoding in the iPhone 8 binary.  The binary agrees with Extractor in every row.}\label{tab:acheong-types}
\begin{tabular}{|>{\raggedright\arraybackslash}p{4.2cm}|>{\raggedright\arraybackslash}p{1.5cm}|>{\raggedright\arraybackslash}p{1.7cm}|>{\raggedright\arraybackslash}p{3.4cm}|>{\raggedright\arraybackslash}p{4.4cm}|}
\hline
\textbf{Message (class)} & \textbf{Field(s)} & \textbf{Published} & \textbf{Extractor} & \textbf{Binary} \\
\hline
\texttt{wloc} request (\texttt{ALSLocation\allowbreak Request}) & 3, 4 & sint32 & int32 & \texttt{Int32Field}, \texttt{i} \\
\hline
\texttt{wloc} request (\texttt{ALSLocation\allowbreak Request}) & 25 & optional & repeated & array ivar, \texttt{Submessage} \\
\hline
LTE cell tower (\texttt{ALSLteCellTower}) & 1--4, 6, 7 & uint32 & int32 & \texttt{Int32Field}, \texttt{i} \\
\hline
\texttt{wloc} location (\texttt{ALSLocation}) & 3--7, 12 & int64 & int32 & \texttt{Int32Field}, \texttt{i} \\
\hline
\texttt{wloc} location (\texttt{ALSLocation}) & 11, 13 & int64 & uint32 & \texttt{Uint32Field}, \texttt{I} \\
\hline
\texttt{pbcwloc} Wi-Fi entry (\texttt{CLPWifiAP\allowbreak Location}) & 9 & int32 & enum & \texttt{i}, has \texttt{scanType\allowbreak AsString:} \\
\hline
\texttt{pbcwloc} location (\texttt{CLPLocation}) & 16, 17 & int32 & bool & \texttt{B} \\
\hline
tile (\texttt{GEOWiFiQuality\allowbreak SearchResult\allowbreak Tile}) & 1 & int64 & uint64 & \texttt{Q}; decodes as varint \\
\hline
tile entry (\texttt{WiFiBSS}) & 4 & pair of int64 & repeated \texttt{WiFiQuality\allowbreak Profile} & array of objects \\
\hline
\end{tabular}
\end{table*}

The first row is an encoding difference rather than a width difference: \texttt{sint32} uses zigzag encoding, so an \texttt{sint32} value of N is encoded as \texttt{2N} and is received as \texttt{2N} by a reader that expects \texttt{int32}. The latitude and longitude fields (\texttt{ALSLocation} 1 and 2) are \texttt{int64} in both. The last row is a difference in wire type: the quality field of a tile entry (field 4 of \texttt{WiFiBSS}) is a repeated message, \texttt{WiFiQualityProfile}, which contains a repeated \texttt{WiFiQualityIndex} of a type and a value. The published descriptor defines it as two \texttt{int64} fields (varint rather than length-delimited), so the real contents are left as unknown fields. The ivar encoding records only that the field is an array of objects, not the class of the elements. Extractor infers the class from the code that reads the field (Section \ref{sec:repeated-fields}), and the 226 captured tiles all decode under it with no unknown fields.

The remaining differences are in the field names. Most are cosmetic and describe the same value (for example \texttt{bssid} and \texttt{macID}, \texttt{wifi\_devices} and \texttt{wirelessAPs}, \texttt{horizontal\_accuracy} and \texttt{accuracy}), so Table \ref{tab:acheong-names} lists only those names that imply a different meaning or that the published descriptor leaves unnamed. Several of the published names for fields of the \texttt{wloc} location message are those of the corresponding \texttt{CLPLocation} fields, but \texttt{ALSLocation} is a different class with only ten data fields, and its fields 7, 11, 12 and 13 are \texttt{confidence}, \texttt{score}, \texttt{reach} and \texttt{infoMask}. The published message also defines fields 8 to 10 and 14 to 31, which do not exist in \texttt{ALSLocation} (the class has ten data ivars and \texttt{writeTo:} writes only tags 1 to 7 and 11 to 13), and none of them occurs in the 2,142 locations in the captured responses. In the \texttt{CoreLocationProtobuf} \texttt{MotionActivity} message, \texttt{writeTo:} writes the \texttt{type} ivar as field 1 and \texttt{confidence} as field 2, the reverse of the published names. The published \texttt{DeviceType} message (operating system, then model) matches the \texttt{ALSMeta} message that the \texttt{wloc} request carries, but \texttt{pbcwloc} uses \texttt{CLPMeta}, which has the product identifier first and the software version second. The last column of Table \ref{tab:acheong-names} gives values observed in the captured traffic. For fields 11 and 12 and for the \texttt{pbcwloc} device information they are consistent with the binary name and not with the published one, and the tile key equals the key requested. The motion activity values do not distinguish the two definitions, so for that message we rely on the binary.

\begin{table*}[htb]
\centering
\footnotesize
\caption{Field names in the published descriptors that imply a different meaning from the iPhone 8 binary, or that are left unnamed.  The binary name is the name of the ivar written for that field number, which Extractor reports unchanged.  The last column gives values observed in the captured traffic.}\label{tab:acheong-names}
\begin{tabular}{|>{\raggedright\arraybackslash}p{3.4cm}|>{\raggedright\arraybackslash}p{1.0cm}|>{\raggedright\arraybackslash}p{3.6cm}|>{\raggedright\arraybackslash}p{2.6cm}|>{\raggedright\arraybackslash}p{1.7cm}|>{\raggedright\arraybackslash}p{4.4cm}|}
\hline
\textbf{Message} & \textbf{Field no.} & \textbf{Published name} & \textbf{Binary name} & \textbf{Meaning} & \textbf{Observed values} \\
\hline
\texttt{wloc} location (\texttt{ALSLocation}) & 4 & \texttt{unknown\_\allowbreak value4} & \texttt{locationType} & unnamed & 0 or 3, 2,001 locations (3 in 550) \\
\hline
 & 7 & \texttt{speed} & \texttt{confidence} & differs & not observed in responses \\
\hline
 & 11 & \texttt{motion\_\allowbreak activity\_\allowbreak type} & \texttt{score} & differs & 1 to 948 (median 63), 2,001 locations \\
\hline
 & 12 & \texttt{motion\_\allowbreak activity\_\allowbreak confidence} & \texttt{reach} & differs & 0 to 70,502 (median 435), 2,001 locations \\
\hline
 & 13 & \texttt{provider} & \texttt{infoMask} & differs & not observed in responses \\
\hline
\texttt{pbcwloc} motion activity (\texttt{CLPMotion\allowbreak Activity}) & 1 & \texttt{confidence} & \texttt{type} & differs (swapped) & 1, 2, 3 \\
\hline
 & 2 & \texttt{activity} & \texttt{confidence} & differs (swapped) & 0, 2 \\
\hline
\texttt{pbcwloc} device information (\texttt{CLPMeta}) & 1 & \texttt{operating\_\allowbreak system} & \texttt{productId} & differs & \texttt{D20AP}, \texttt{D53gAP} \\
\hline
 & 2 & \texttt{model} & \texttt{softwareVersion} & differs & \texttt{iPhone OS16.7.11/\allowbreak 20H360}, \texttt{iPhone OS16.1.1/\allowbreak 20B101} \\
\hline
tile (\texttt{GEOWiFiQuality\allowbreak SearchResult\allowbreak Tile}) & 1 & \texttt{unknown1} & \texttt{tileKey} & unnamed & equals the request \texttt{X-tilekey} header in 226 of 226 tiles \\
\hline
\end{tabular}
\end{table*}

The Extractor results are for iOS 16.7.11, whereas the published descriptors are for later iOS versions: the repository's code identifies itself to Apple as iOS 17.5 and its report analyses iOS 18.6.2.

\subsubsection{CloudKit Protobuf Descriptors}\label{cloudkit-protobuf-descriptors}

CloudKit is the protocol used by iCloud and iMessage to store and synchronise data. The iMessage client library rustpush \citep{rustpush} contains a protobuf descriptor file for it, \texttt{cloudkit.proto}, with 153 messages. Field names such as \texttt{unk2} and comments such as \texttt{always\ 0} indicate that it was written by hand from observed traffic. We exclude three messages of the iCloud content-storage framework MMCS, which use protobuf-c, an encoding that Extractor does not support. We compare the remaining 150 messages with the descriptors that Extractor recovers from the iPhone 8, counting the generic \texttt{SignedInfo} as the three messages it stands for in the binary (152 in all). 142 of the 152 pair with an Extractor message and ten have no counterpart. Of the 142 pairs, 56 pair under a name that differs from the rustpush name (for example rustpush \texttt{AssetUploadData} is \texttt{CKDPAsset} in the binary), which we identified from similar names or from matching field numbers and types. Of the ten unmatched, two (\texttt{FileToken} and \texttt{FileTokens}) are defined by rustpush but are unused, five are empty in rustpush or were added after iOS 16 and are in iOS 27, where Extractor recovers them and their fields, and three are in no binary that we examined.

The 142 pairs share 592 fields. The field numbers and wire types agree wherever both descriptors define a field, with one exception: \texttt{SupportsSOS}, which rustpush defines with a boolean as field 2 and the binary with an enumeration as field 1. Four differ in repeated or singular label, for example the published descriptor has a single \texttt{asset} in \texttt{AssetUpload} where the binary has an array. 387 of the 592 agree in number, name, type and label. 147 differ in name and 74 in declared type. Table \ref{tab:rustpush-types} lists the type differences with the evidence. In every case the binary agrees with Extractor.

\begin{table*}[htb]
\centering
\footnotesize
\caption{Differences in declared field type between the rustpush descriptors and Extractor for the 142 paired CloudKit messages, with the evidence in the iPhone 8 binaries.  The binary agrees with Extractor in all 74 cases.}\label{tab:rustpush-types}
\begin{tabular}{|>{\raggedright\arraybackslash}p{4.2cm}|>{\raggedright\arraybackslash}p{2.0cm}|>{\raggedright\arraybackslash}p{2.0cm}|p{0.8cm}|>{\raggedright\arraybackslash}p{6.4cm}|}
\hline
\textbf{Fields} & \textbf{rustpush} & \textbf{Extractor} & \textbf{n} & \textbf{Binary} \\
\hline
enumerations, e.g.\ \texttt{RequestOperation.Header} 35, \texttt{Subscription} 2, \texttt{ShareInfo} 3, \texttt{ResetRequest} 1 & int32 or uint32 & enum & 30 & the class has a \texttt{<field>AsString:} method, or the Swift type is an enumeration \\
\hline
nested messages, e.g.\ \texttt{SubscriptionNotification} 1, \texttt{DispositionType} 1--5 & bytes & message & 18 & ivars of a message class (\texttt{@"CKDPSubscriptionNotificationAlert"}), or arrays written by the routine that writes message fields \\
\hline
\texttt{Asset} 4, 14, 18 & uint64 & int64 & 3 & ivar \texttt{q} (signed) \\
\hline
\texttt{TlkShare} 2, 3, 5, 10 & uint64 or uint32 & int64 & 4 & Swift type \texttt{Int64} \\
\hline
\texttt{EncryptedAttributes} 3--10, 14, 15 & uint64, uint32 or int32 & int32, int64 or uint32 & 10 & ivars \texttt{i}, \texttt{q} or \texttt{I}, as recovered by Extractor (all 10) \\
\hline
\texttt{AssetUploadData} 4, \texttt{AssetInfo} 3, \texttt{MMCSHeaders} 2 & uint32, uint64 or int32 & int64 or uint64 & 3 & ivars \texttt{q}, \texttt{q} and \texttt{Q} \\
\hline
\texttt{RequestOperation.Header} 29, 34, \texttt{ZoneDeleteRequest} 2 & int32 or uint32 & bool & 3 & ivar \texttt{B} \\
\hline
\texttt{SubscriptionNotification} 3, \texttt{EncryptedAttributes} 11 & bytes & string & 2 & array written by the string routine; ivar \texttt{NSString} \\
\hline
\texttt{RecordSaveRequest} 3 & repeated string & repeated message & 1 & array written by the message routine, not the string routine \\
\hline
\end{tabular}
\end{table*}

Most of these are places where the published descriptor is less specific than Extractor (an integer where the class has an enumeration, or \texttt{bytes} where the field is a nested message or a string) or where it uses an integer of a different width or sign. These decode identically on the wire for values that fit both types. The last row is an error in the published descriptor, since a repeated string cannot decode the field's contents.

Of the 147 name differences, 7 are \texttt{unk} placeholders that Extractor names (for example \texttt{unk1} is \texttt{deviceSoftwareIsAppleInternal}), four are spelling errors in the published descriptor (\texttt{lattitude}, \texttt{evaulationType}, \texttt{isAnnonymousInvitedParticipant}, \texttt{annonymousPublicAccess}), and the rest are different wordings. For \texttt{OTBottle}, the keys that the published descriptor names \texttt{escrowedSigningKey}, \texttt{escrowedEncryptionKey}, \texttt{peerSigningKey} and \texttt{peerEncryptionKey} are named with the suffix \texttt{SPKI} in the binary (the standard X.509 SubjectPublicKeyInfo encoding of a public key), for example \texttt{escrowedSigningSPKI}, and \texttt{ciphertext} is named \texttt{contents}.

The descriptors differ in coverage. Extractor recovers 184 fields that rustpush does not define. Rustpush defines 31 fields that are not in the iOS 16.7.11 binaries. We repeated the comparison on iOS 27.0.1, where 27 of the 31 appear, so they were added after iOS 16. The other four (\texttt{operationCost} and fields 3, 7 and 12 of \texttt{Asset}) are in none of the binaries we searched.

\subsubsection{iMessage Protobuf Descriptors}\label{imessage-protobuf-descriptors}

The iMessage client library rustpush \citep{rustpush} defines the outer envelope of an iMessage in a small descriptor file, \texttt{ids.proto}, with five messages (one nested). We compare it with the descriptors that Extractor recovers from the \texttt{MessageProtection} framework of the iPhone 8, where the corresponding classes have the prefix \texttt{NGMPB}. Four of the five messages pair with an Extractor message (\texttt{PreKeyData} is \texttt{NGMPBPublicDevicePrekey}, \texttt{OuterMessage} is \texttt{NGMPBOuterMessage}, \texttt{InnerMessage} is \texttt{NGMPBInnerMessage} and the nested \texttt{NgmPublicIdentity} is \texttt{NGMPBPublicDeviceIdentity}). The fifth, \texttt{KtLoggableData}, has no counterpart: the iPhone 8 code receives it as an opaque \texttt{NSData} blob and passes it to a \texttt{KTLoggableData} class that stores it as bytes.

The four pairs share 12 fields (Table \ref{tab:ids-fields}). The field numbers, wire types and declared types agree in every case, including the \texttt{uint32} counter and the \texttt{double} timestamp, and the \texttt{writeTo:} code of the iPhone 8 classes writes exactly these field numbers. The only differences are in 6 field names. The binary names describe the role of the field (for example \texttt{encryptedPayload}, \texttt{ephemeralPubKey} and \texttt{keyValidator}), where rustpush uses \texttt{payload}, \texttt{key} and \texttt{validator}. This is the one set of published descriptors that we compare with no difference in type.

\begin{table*}[htb]
\centering
\footnotesize
\caption{Fields of the iMessage envelope in the rustpush descriptor and in the iPhone 8 binary.  Field numbers and types agree for all 12 fields, and only the names differ.}\label{tab:ids-fields}
\begin{tabular}{|>{\raggedright\arraybackslash}p{4.4cm}|p{1.3cm}|>{\raggedright\arraybackslash}p{3.2cm}|>{\raggedright\arraybackslash}p{3.6cm}|p{2.2cm}|}
\hline
\textbf{Binary class (rustpush message)} & \textbf{Field} & \textbf{rustpush name} & \textbf{Binary name} & \textbf{Type} \\
\hline
\texttt{NGMPBPublicDevicePrekey} (\texttt{PreKeyData}) & 1 & \texttt{key} & \texttt{prekey} & bytes \\
 & 2 & \texttt{signature} & \texttt{prekeySignature} & bytes \\
 & 3 & \texttt{timestamp} & \texttt{timestamp} & double \\
\hline
\texttt{NGMPBPublicDeviceIdentity} (\texttt{NgmPublicIdentity}) & 1 & \texttt{publicKey} & \texttt{signingKey} & bytes \\
\hline
\texttt{NGMPBInnerMessage} (\texttt{InnerMessage}) & 1 & \texttt{message} & \texttt{message} & bytes \\
 & 2 & \texttt{counter} & \texttt{counter} & uint32 \\
 & 3 & \texttt{ktGossipData} & \texttt{ktGossipData} & bytes \\
 & 99 & \texttt{debugInfo} & \texttt{debugInfo} & bytes \\
\hline
\texttt{NGMPBOuterMessage} (\texttt{OuterMessage}) & 1 & \texttt{payload} & \texttt{encryptedPayload} & bytes \\
 & 2 & \texttt{key} & \texttt{ephemeralPubKey} & bytes \\
 & 3 & \texttt{signature} & \texttt{signature} & bytes \\
 & 99 & \texttt{validator} & \texttt{keyValidator} & bytes \\
\hline
\end{tabular}
\end{table*}

\subsection{Access Key And Maps-Auth-Token Generation}\label{sec:access-key}

Resource requests to \texttt{gspe21-ssl.ls.apple.com} carry an \texttt{accessKey} query parameter, and tile requests to \nolinkurl{gspe19-ssl.ls.apple.com/tile.vf} carry a \texttt{Maps-Auth-Token} header. The handset computes both from a salt string hard-coded in the GeoServices framework and the \texttt{authToken} value \(T\) that it receives from \nolinkurl{gspe35-ssl.ls.apple.com/geo_manifest/dynamic/config}. The \texttt{accessKey} also depends on the path of the requested resource, the current time and a random nonce, and the \texttt{Maps-Auth-Token} on the version of the requested map style. None of these inputs is specific to the handset or the user.

In tests we found that the same \(T\) is sent to different iPhones, both when they have the same IP address and when they have different ones, and that different iPhones send the same \texttt{Maps-Auth-Token} values. Neither value therefore acts as an individual device identifier.

\subsection{\texorpdfstring{\texttt{X-tilekey} Header Decoding}{X-tilekey Header Decoding}}\label{x-tilekey-header-decoding}

A Gloria tile ID is a 64-bit unsigned integer \(E\) that encodes a Web~Mercator tile's grid coordinates \((x,\, y)\) and zoom level \(z\) in a single value. The three decoding steps are described below.

\subsubsection{Zoom Recovery}\label{zoom-recovery}

During encoding, a \emph{sentinel bit} is placed at bit~position~\(2z\): \[E = m \;\big|\; 4^{z}\] where \(m\) is the Morton code and \(4^{z} = 2^{2z}\). Because the Morton code for a zoom-\(z\) tile uses only bits \(0\ldots 2z-1\), the most-significant set bit of \(E\) is always exactly bit \(2z\).

Let \(\nu = \lfloor\log_{2} E\rfloor\) be the zero-indexed position of the most-significant set bit of \(E\). Then \[{z = \left\lfloor \frac{\nu}{2} \right\rfloor}\]

\subsubsection{Sentinel Removal}\label{sentinel-removal}

Since \(4^{z}\) is now known, the raw Morton code is recovered by XOR-clearing the sentinel bit: \[{m = E \;\oplus\; 4^{z}}\]

No masking of lower bits is needed: the Morton code carries no information at or above position \(2z\) by construction.

\subsubsection{Morton Code De-Interleaving}\label{morton-code-de-interleaving}

The Morton code is formed by interleaving the bits of \(x\) and \(y\): \[m = \mathrm{spread}(x) \;\big|\; \bigl(\mathrm{spread}(y) \ll 1\bigr)\] where \(\mathrm{spread}\) expands a 32-bit integer so each bit occupies an even-indexed position (0, 2, 4, \(\ldots\)).

Decoding uses the inverse \textbf{compact} operation, which gathers the even-indexed bits of a 64-bit word back into the contiguous low-order bits of a 32-bit result: \[{x = \mathrm{compact}(m), \qquad y = \mathrm{compact}(m \gg 1)}\]

\subsubsection{The Compact Operation}\label{the-compact-operation}

Compact proceeds as an initial mask followed by five mask-and-fold stages. Define the 64-bit masks \[M_k = \frac{2^{64}-1}{2^{2^{k+1}}-1}\left(2^{2^k}-1\right), \quad k = 0,\ldots,5,\] in which each block of \(2^{k+1}\) bits has its low \(2^k\) bits set, or, in hexadecimal: \[\begin{aligned}
  M_0 &= \texttt{0x5555555555555555} \quad (\text{even bits only}) \\
  M_1 &= \texttt{0x3333333333333333} \quad (\ldots 0011\,0011_2) \\
  M_2 &= \texttt{0x0F0F0F0F0F0F0F0F} \quad (\ldots 0000\,1111_2) \\
  M_3 &= \texttt{0x00FF00FF00FF00FF} \\
  M_4 &= \texttt{0x0000FFFF0000FFFF} \\
  M_5 &= \texttt{0x00000000FFFFFFFF}
\end{aligned}\]

For input \(n\) (either \(m\) or \(m \gg 1\)): \begin{align}
  n &\leftarrow n \;\mathbin{\&}\; M_0 \label{eq:c0}\\
  n &\leftarrow (n \mid n \gg 1)  \;\mathbin{\&}\; M_1 \label{eq:c1}\\
  n &\leftarrow (n \mid n \gg 2)  \;\mathbin{\&}\; M_2 \label{eq:c2}\\
  n &\leftarrow (n \mid n \gg 4)  \;\mathbin{\&}\; M_3 \label{eq:c3}\\
  n &\leftarrow (n \mid n \gg 8)  \;\mathbin{\&}\; M_4 \label{eq:c4}\\
  n &\leftarrow (n \mid n \gg 16) \;\mathbin{\&}\; M_5 \label{eq:c5}
\end{align}

Step~\eqref{eq:c0} isolates the target bits (even-indexed for \(x\), odd-indexed already shifted to even for \(y\)). Each subsequent step folds adjacent pairs of already-gathered bits downward by a doubling shift, progressively packing them into the low 32~bits. The final mask \(M_5\) discards the upper 32~bits, leaving a clean 32-bit result.

\subsubsection{Result}\label{result}

The decoded tile coordinates \((x,\, y,\, z)\) are:

\begin{itemize}
\item
  \(z\): the Web~Mercator zoom level; the tile space is a \(2^z \times 2^z\) grid.
\item
  \(x \in [0,\, 2^{z})\): tile column, increasing west to east.
\item
  \(y \in [0,\, 2^{z})\): tile row, increasing north to south (standard Web~Mercator convention).
\end{itemize}

\begin{table}[htb]
\centering
\footnotesize
\begin{tabular}{|p{3cm} | p{5cm} |}
\hline
\textbf{Onboarding screen} & \textbf{Option selected} \\
\hline
Language & "English", "Ireland"\\
Quick Start & "Set Up Manually"\\
Written and Spoken Languages & "Continue"\\
Choose a Wi-Fi Network & "Continue without wifi", "Continue"\tablefootnote{An "Unable to activate" screen is displayed.  We click "Other options", "Use Mac or PC", then connect the iPhone to a MacBook using a USB cable.  On the MacBook we open Apple Music, click the iPhone icon, then "Open in Finder".  A popup says "Activating" and, after a delay, the iPhone moves on to the next screen.  We then disconnect the USB cable.}\\
Data \& Privacy &"Continue"\\
Touch ID & "Set up Touch ID later"\\
Create a Passcode & "Passcode options", "Don't use passcode"\\
Apps \& Data & "Don't Transfer Apps \& Data" \\
Apple ID &"Forgot password or don't have an Apple ID", "Set up later in Settings"\\
Terms and Conditions & "Agree"\\
Update your iPhone Automatically & "Only Download Automatically"\\
iMessage \& FaceTime & "Not now"\\
Location Services &  "Disable Location Services" \\
Siri & "Set Up Later in Settings"\\
Screen Time & "Set Up Later in Settings"\\
iPhone Analytics &"Don't Share" \\
Appearance & "Continue"\\
Display Zoom &"Continue"\\
Welcome to iPhone &"Get Started"\\
\hline
\end{tabular}
\caption{Settings selected during handset onboarding following a factory reset.}\label{tab:settings}
\end{table}

\subsection{\texorpdfstring{Connections Made By \texttt{geod} And \texttt{locationd}}{Connections Made By geod And locationd}}\label{connections-made-by-geod-and-locationd}

\begin{table*}[htb]
\begin{threeparttable}[b]
\scriptsize
\centering
\caption{Network connections made by the iOS \texttt{geod} process.}
 \label{tab:geod-connections}
\begin{tabular}{@{}
  p{3.1cm}   %
  p{1.7cm}   %
  p{3.7cm}   %
  p{3.6cm}   %
  p{1.7cm}   %
  p{0.9cm}   %
@{}}
\toprule
\textbf{Endpoint} &
\textbf{Purpose} &
\textbf{Request content} &
\textbf{Response content} &
\textbf{Loc. sent} &
\textbf{When loc. off?} \\
\midrule

\nolinkurl{configuration.ls.apple.com/config/defaults}
& Settings
& URL params:  {OS version}, {hardware}
& XML plist: feature flags,
  tile-type switches, endpoint URLs, per-country feature maps
& No & $\checkmark$ \\
\hline

\nolinkurl{gspe35-ssl.ls.apple.com/geo_manifest/dynamic/config}
& Settings
& URL params:   {country code}, {OS version}, {hardware}
& Protobuf: \texttt{geod} resources, incl.\ map tiles, fonts, icons, textures available, endpoint URLs
& Country code & $\checkmark$ \\
\hline

\nolinkurl{gspe1-ssl.ls.apple.com/pep/gcc}
& IP country lookup
& Standard HTTP headers only; no body
& Plain text: Country code
& No & $-$ \\
\hline

\nolinkurl{gsp-ssl.ls.apple.com/ab.arpc}
& A/B experiments
& ARPC protobuf: session GUID, locale, OS version,
  country code
& Protobuf: bucket ID for namespaces ({parsec},
  {Siri}, {RAP}, {mapsAb})
 & Country code 
& $\checkmark$ \\
\hline

\nolinkurl{gsp64-ssl.ls.apple.com/hvr/v3/use}
& Telemetry
& ARPC protobuf: batched per-request log records
  with app ID, HW/OS, connectivity type, carrier, country code,
  latency, HTTP status, server IP:port, tile-set ID
& Empty ARPC acknowledgement (0-byte payload)
& Carrier; country code & $\checkmark$ \\
\hline

\nolinkurl{gsp9-ssl.apple.com/hvr/v2/rtloc}
& Real-time traffic probe
& ARPC protobuf: sequence of timestamped records with handset 
GPS location, accuracy, handset motion type
& Empty ARPC acknowledgement (0-byte payload)
& Precise (GPS, $\approx$5\,m accuracy) & $-$ \\
\hline

\nolinkurl{gsp-ssl.ls.apple.com/dispatcher.arpc}
& Rev. geocoding, geocoding
& ARPC protobuf: caller (com.apple.weather.widget, analyticsd, com.apple.networkserviceproxy,
  com.apple.kbd or com.apple.corelocation.CoreLocationTemporaryPreciseAuthPromptPlugin), HW/OS, locale,
  coarse coordinates or none, place-data
  & Protobuf: reverse-geocoded street address, bounding polygons, place-data
& Coarse or none\tnote{1} & $\checkmark$ \\
\hline

\nolinkurl{gsp57-ssl-revgeo.ls.apple.com/dispatcher.arpc}
& Rev. geocoding
& ARPC protobuf: caller com.apple.CoreRoutine.helperservice, HW/OS, locale,
  GPS coordinates, place-data
& Protobuf: reverse-geocoded street address, bounding polygons, place-data
& Precise (full-precision coordinates) & $-$ \\
\hline

\nolinkurl{gsp57-ssl-locus.ls.apple.com/dispatcher.arpc}
& Wi-Fi fingerprint lookup
& ARPC protobuf: caller com.apple.CoreRoutine.helperservice, HW/OS, locale, session ID,
  timestamped locations, MAC address and signal strength of nearby Wi-Fi APs, visit entry/exit times
& Status only: no result
& Precise (reported accuracy 5.6 to 30 m) & $-$ \\
\hline

\nolinkurl{gsp85-ssl.ls.apple.com/wifi_request}
& Wi-Fi quality lookup
& ARPC protobuf: BSSID of currently associated
  AP, locale, country code
& Protobuf: quality status for submitted BSSID
& BSSID (current Wi-Fi AP) & $\checkmark$ \\
\hline

\nolinkurl{gspe85-ssl.ls.apple.com/wifi_request_tile}
& Wi-Fi quality tile download
& {X-tilekey} header: Gloria tile key\tnote{2}
& Protobuf tile: per-AP records with packed BSSID (48-bit int),
  lat/lon, popularity score, auth type, public-network flag
& Coarse (tile area\tnote{3}) & $-$ \\
\hline

\nolinkurl{gspe19-ssl.ls.apple.com/tile.vf}
& Map resource download
& URL params or {maps-tile-*} headers: 
  Mercator tile coords x, y, z\tnote{4} 
  & VMP4 file~\cite{spangenberger20,vmp4dump}\tnote{5}: map features, POIs etc.
& Coarse (tile area\tnote{6})
& $\checkmark$  \\
\hline

\nolinkurl{gspe79-ssl.ls.apple.com/65/v1/zoom8/<key>.tar.gz}
& Coarse location tile
& URL path: Gloria tile key\tnote{2}
 & Coarse location grid cells
& Coarse (tile area\tnote{7}) & $-$ \\
\hline

\nolinkurl{gspe79-ssl.ls.apple.com/72/v2/72/<key>}
& Network interface selection (Wi-Fi/LTE/5G)
& URL path: Gloria tile key\tnote{2}
& Tiles with network data: downlink throughput,
  interface mask, quality percentile, tower hash
& Coarse (tile area\tnote{3}) & $-$ \\
\hline

\nolinkurl{gspe21-ssl.ls.apple.com/stylesheet/*.styl}
& VectorKit stylesheet download
& URL param; version in filename
& VectorKit rendering rules (colours, stroke
  widths, label fonts, icon zoom ranges, POI busyness hooks)
& No & $\checkmark$ \\
\hline

\nolinkurl{gspe21-ssl.ls.apple.com}
(\nolinkurl{/xml/}, \nolinkurl{/other/},
\nolinkurl{/icon/}, \nolinkurl{/config/})
& Map resource downloads
& URL param; version in filename
& JSON/plist/protobuf/XML: POI type maps, country feature lists,
  localisation data, altitude refs, publisher attribution,
  EV/vehicle info; C3H/C3B 3D tiles;
  icons: transit, POI, localised
& No & $\checkmark$ \\
\hline

\nolinkurl{mesu.apple.com/assets/.../GeoPolygonDataAssets.xml}
& Settings
& OS version
& XML manifest of available polygon data
& No & $\checkmark$ \\

\bottomrule
\end{tabular}
\begin{tablenotes}
\item [1] By caller: city-level (one fixed point several kilometres from the handset) for com.apple.weather.widget; coordinates rounded to 1 to 2 decimal places for analyticsd; reported accuracy of about 3.7\,km for com.apple.networkserviceproxy; postal district, no coordinates, for com.apple.kbd; IP address only for com.apple.corelocation.CoreLocationTemporaryPreciseAuthPromptPlugin.
\item [2] Encoding lat/lon/zoom $\approx$1--2\,km at zoom~13;  e.g.\ 81607936 $\to$ 53.32°N, 6.31°W
\item [3] Maximum tile zoom observed is 13, corresponding to a tile size of $\approx$0.05$^\circ$ or $\approx$5\,km. 
\item [4] Maps to lat/lon/zoom, e.g.\ x=1976, y=1328, z=12 (zoom) $\to$ 52.1°N, 6.33°W
\item [5] All successful \texttt{tile.vf} responses in our iPhone 8 captures begin with the bytes \texttt{VMP4} (after optional gzip), followed by a table of typed sections, some of which are zlib compressed, as in the on-device map cache files described in \cite{spangenberger20}.  We do not decode the contents.
\item [6] Maximum zoom level observed is 15, corresponding to a tile size of $\approx$0.01$^\circ$ or $\approx$1\,km.
\item [7] Zoom 8, corresponding to a tile size of $\approx$1.4$^\circ$ or $\approx$150\,km.
\end{tablenotes}
\end{threeparttable}
\end{table*}

\begin{table*}[!ht]
\begin{threeparttable}[b]
\centering
\caption{
  HTTPS connections made by the iOS \texttt{locationd} process.
}
\label{tab:locationd-connections}
\scriptsize
\begin{tabular}{@{}
    p{4.5cm}   %
    p{2.3cm}   %
    p{3.6cm}   %
    p{3.6cm}   %
    p{1.7cm}   %
    p{0.9cm}   %
  @{}}
\toprule
\textbf{Endpoint} &
\textbf{Purpose} &
\textbf{Request content} &
\textbf{Response content} &
\textbf{Loc.\ sent} &
\textbf{When loc. off?} \\
\midrule

\texttt{gs-loc.apple.com} \newline
\texttt{/clls/wloc}
&
Wi-Fi/cell location lookup
&
Request contains a cell tower identity (MCC, MNC, cell ID) or a Wi-Fi MAC address (in one request, the handset lat/lon).
&
Response is a list of cell towers or Wi-Fi MAC addresses and their lat/lon positions.
&
Precise (Wi-Fi/cell tower coverage area; unrounded lat/lon in one request)
&
$-$
\\
\hline

\texttt{gsp10-ssl.apple.com/hcy/pbcwloc}
&
Cell/Wi-Fi observation upload
&
Cell tower (MCC, MNC, TAC, cell ID, signal strength) or Wi-Fi (MAC address, channel, signal strength) records, each with the handset location and motion state
&
Acknowledgement only
&
Precise (10--400\,m)
&
$-$
\\
\hline

\texttt{gsp10-ssl.apple.com/pds/pd}
&
GPS fix upload
&
GPS fixes (lat/lon, altitude, accuracy, speed, course, motion state), app bundle ID, trip ID
&
Acknowledgement only
&
Precise (GPS, $\approx$5\,m)
&
$-$
\\
\hline

\texttt{gsp10-ssl.apple.com/hvr/trc}
&
Sensor trace upload
&
Pressure, temperature, locations with GNSS satellite data, compass attitude, pedometer, motion state, app bundle ID
&
Empty
&
Precise (GPS and network fixes)
&
$-$
\\
\hline

\texttt{gsp10-ssl.apple.com/psr}
&
Pressure upload
&
Barometric pressure reading and timestamp, with the GPS fix at the time of the reading (lat/lon, altitude, accuracy)
&
Empty
&
Precise (GPS, $\approx$5\,m)
&
$-$
\\
\hline

\texttt{cl2.apple.com/1/v1/} \newline
\texttt{\{lat0\}/\{lon0\}/\{lat\}\_\{lon\}.gz} \newline
&
Wi-Fi tile download
&
Tile SW corner in URL path\tnote{1}
&
A tile containing Wi-Fi MAC addresses and lat/lon positions
&
Coarse (tile size, $\approx$5\,km) 
&
$-$
\\
\hline

\texttt{cl2.apple.com/3/v1/} \newline
\texttt{\{lat0\}/\{lon0\}/} \newline
\texttt{\{lat\}\_\{lon\}\_\{sec\_key\}.gz} \newline
&
GSM cellular tile download
&
Tile SW corner in URL path\tnote{1}, \texttt{sec\_key}
$=\!(\mathrm{MCC}\!\ll\!16)|\mathrm{MNC}$ encodes mobile carrier
&
GSM cell identifiers, locations and coverage footprints for the specified carrier network
&
Coarse (tile size, $\approx$130\,km)
&
$-$
\\
\hline

\texttt{cl2.apple.com/4/v1/} \newline
\texttt{\{lat0\}/\{lon0\}/} \newline
\texttt{\{lat\}\_\{lon\}\_\{sec\_key\}.gz} \newline
&
LTE cellular tile download
&
Tile SW corner in URL path\tnote{1}, \texttt{sec\_key}
$=\!(\mathrm{MCC}\!\ll\!16)|\mathrm{MNC}$ encodes mobile carrier
&
LTE cell identifiers (TAC and Global CellID) and their locations for the specified carrier network
&
Coarse (tile size, $\approx$130\,km)
&
$-$
\\
\hline

\texttt{cl3.apple.com/2/v1/} \newline
\texttt{\{lat0\}/\{lon0\}/\{lat\}\_\{lon\}.gz} \newline
&
High-Resolution Signal Environment (HRSE) tile download
&
Tile SW corner in URL path\tnote{1}
&
RF signal environment class\tnote{2} at fine resolution over the tile area
&
Coarse  (tile size, $\approx$5\,km) 
&
$-$
\\
\hline

\texttt{cl3.apple.com/3/v2/} \newline
\texttt{\{lat0\}/\{lon0\}/\{lat\}\_\{lon\}.gz} \newline
&
Digital elevation model (DEM) tile download
&
Tile SW corner in URL path\tnote{1}
&
Altitudes on a grid across the tile (likely used for GPS altitude
correction and barometric fusion)
&
Coarse (tile size, $\approx$7\,km)
&
$-$
\\
\hline

\texttt{cl2.apple.com/\{6,7,8,11\}/v1/} \newline
\texttt{0/0/0000000\_0000000.gz}
&
Global cellular tile download
&
Path \texttt{0/0/0000000\_0000000} is a single tile covering all coordinates, so no location information is conveyed
&
Global cellular tile: /6 $\to$ list of 87K GSM MCC/MNC/CelIID entries and corresponding lat/lon locations, /7 $\to$ list of CDMA Base Station IDs, /8 $\to$ list of LTE Global Cell IDs (eNB-ID+antenna index), /11 $\to$ list of 5G NR Cell IDs.
&
$-$
&
$-$
\\
\hline

\texttt{cl3.apple.com:} \newline
\texttt{\quad cldavl.avl.gz} \newline
\texttt{\quad clhrseavl.avl.gz} \newline
\texttt{\quad clPressureavl.avl.gz}
&
Tile availability indexes for DEM, HRSE, and barometric-pressure tile sets
&
-
&
Availability bitmaps indicating which geographic tiles have server-side data
&
$-$
&
$-$
\\
\hline

\texttt{cl3.apple.com:} \newline
\texttt{\quad clsef.bin.gz}
&
Global coarse signal-environment map
&
-
&
Global raster map of RF signal environment\tnote{2}, covering the Earth at 0.05$^\circ$ resolution ($\approx$5.5\,km per cell at the equator).
&
$-$
&
$-$
\\
\hline

\texttt{cl4.apple.com/2/m2/a.1//} \newline
\texttt{availability.pb.gz}
&
Indoor positioning venue availability
&
-
&
List of venue bounding boxes (NW/SE lat/lon pairs) for regions
where indoor map tiles exist
&
$-$
&
$-$
\\
\hline

\texttt{iphone-ld.apple.com} \newline
\texttt{/xtra/xtra3grej.bin} \newline
\texttt{/xtra/xtra3int.bin}
&
GNSS orbit assistance data (Qualcomm XTRA, proxied by Apple)
&
-
&
Satellite ephemeris/almanac data\tnote{3}
&
$-$
&
\checkmark
\\
\hline

\texttt{configuration.apple.com} \newline
\texttt{/.../bt/bcwv.plist} \newline
\texttt{/.../bt/bcw.plist.gz}
&
Bluetooth hands-free vehicle allowlist
&
-
&
\texttt{bcwv.plist}: XML plist containing only a version integer (15);
\texttt{bcw.plist.gz}: gzip plist with BT device-name patterns and
MAC-address prefix allowlist
&
$-$
&
\checkmark
\\

\bottomrule
\end{tabular}
\begin{tablenotes}
\item [1] lat/lon in URL are encoded as lat=(latitude + 90) $\times$ 10000, lon=(longitude + 180) $\times$ 10000, lat0/lon0 are the first three digits of lat/lon.
\item [2] Signal environment types are: Rural, Urban, DenseUrban, DenseUrbanCanyon, Marine, Foliage
\item [3] Current satellite location data plus estimated locations for the next 7 days and signal information, see \nolinkurl{https://wwws.nightwatchcybersecurity.com/2016/12/05/cve-2016-5341/}
\end{tablenotes}
\end{threeparttable}
\end{table*}

\subsection{\texorpdfstring{Third-Party Claims About \texttt{/hvr/v3/use}}{Third-Party Claims About /hvr/v3/use}}\label{third-party-claims-about-hvrv3use}

The technical report accompanying \citep{acheong08} states that \texttt{gsp64-ssl.ls.apple.com/hvr/v3/use} receives opaque binary data that includes the user's home location, the open apps, and IP addresses and ports, and that some of the requests are encrypted. The report does not decode the payloads. Our decoded records (Section \ref{sec:telemetry}) account for most of these observations. The ``open apps'' correspond to the identifier of the app or daemon that triggered each \texttt{geod} request (for example \texttt{com.apple.wifi.ThreeBarsXPCService}), which is not a list of the apps that are open. The IP addresses and ports are those of the Apple server that the request was sent to (for example \texttt{17.253.77.204:443}, in 15 records), and not the handset's connections. The requests that appear encrypted are ARPC bodies compressed with LZ4, which have high entropy until they are decompressed. We did not observe a home location. The Maps user record has the fields \texttt{homeCountryCode} and \texttt{homeMetroRegion}, but both read ``UNKNOWN'' because our test user has no home set, and the fields indicate that a populated value would be a country and metro region and not a street address.

\subsection{Handset Onboarding Settings}\label{handset-onboarding-settings}

Table \ref{tab:settings} lists the settings that the test user selected during onboarding after the factory reset.

\subsection{Further Figures}\label{further-figures}

\begin{figure}[htb]
\centering
\begin{subfigure}[b]{0.45\columnwidth}
\includegraphics[width=\columnwidth]{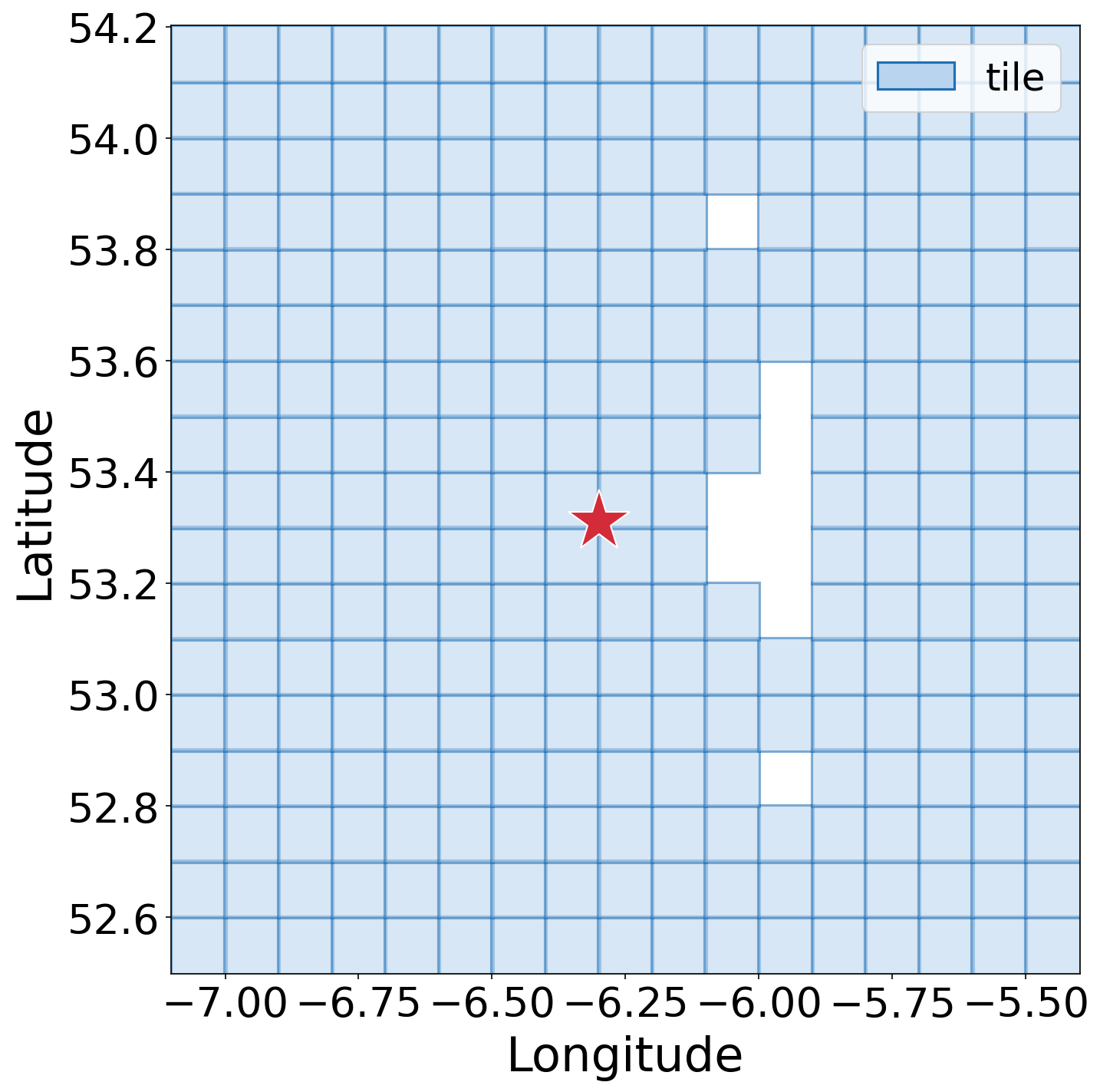}
\subcaption{DEM tile}
\end{subfigure}
\begin{subfigure}[b]{0.45\columnwidth}
\includegraphics[width=\columnwidth]{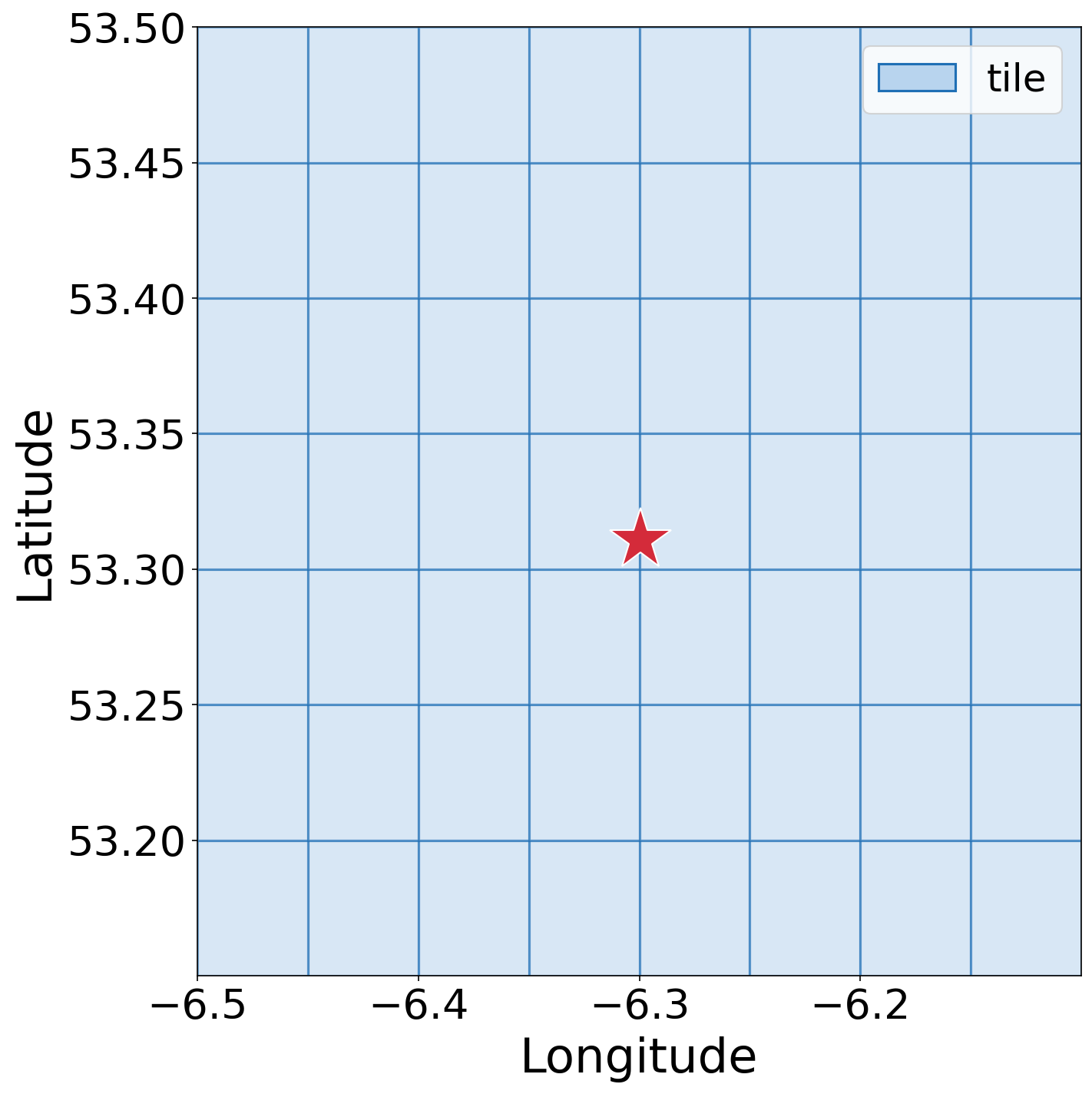}
\subcaption{Wi-Fi tile}
\end{subfigure}

\vspace{1mm}
\begin{subfigure}[b]{0.45\columnwidth}
\includegraphics[width=\columnwidth]{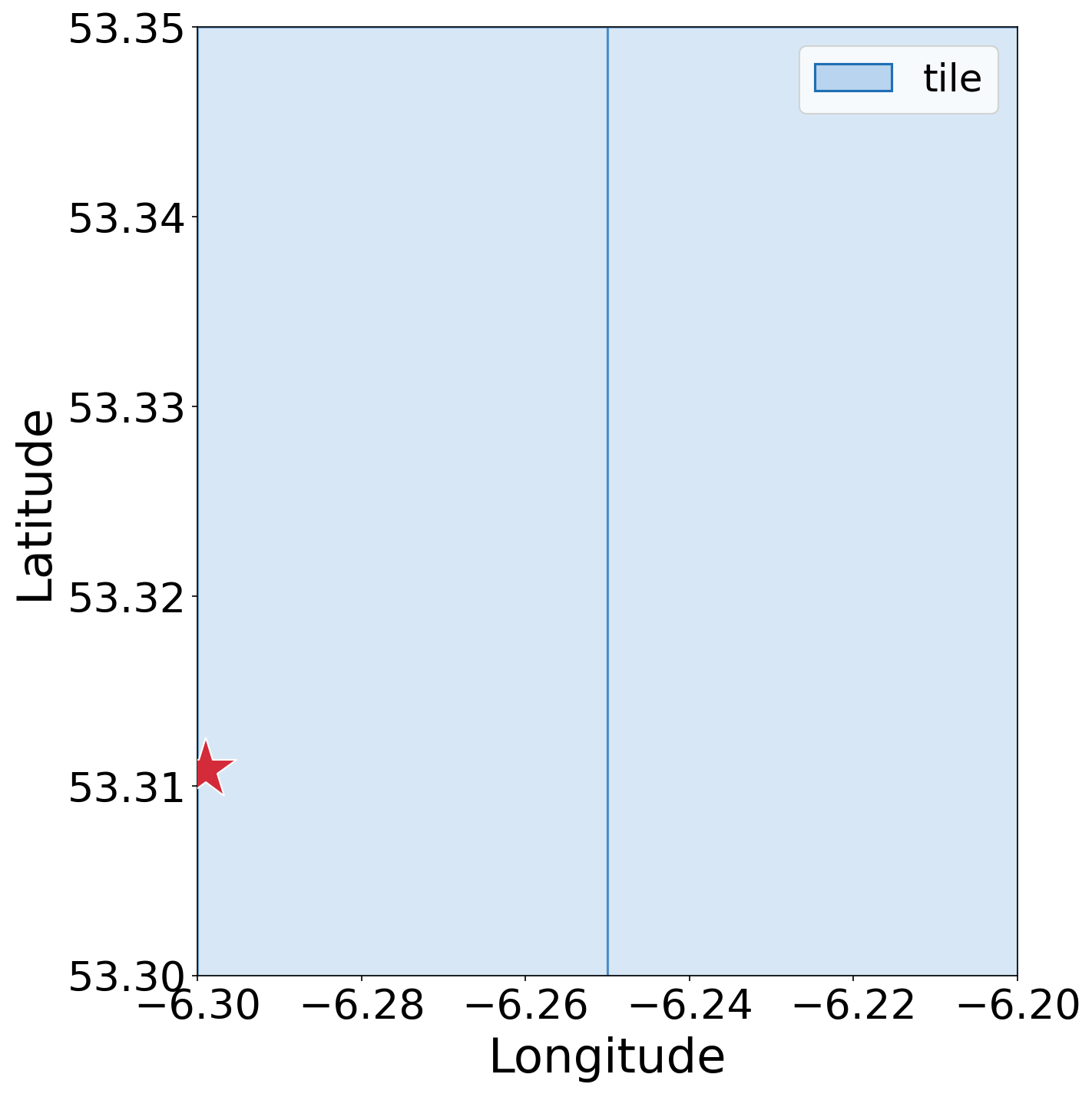}
\subcaption{HRSE tile}
\end{subfigure}

\caption{Visualisation of smaller tiles ($\approx$3--11\,km across) downloaded by \texttt{locationd}.  The star in each plot marks the handset location.}\label{fig:locd_tiles}
\end{figure}

Figure \ref{fig:locd_tiles} shows the smaller tiles (about 3--11 km across) that \texttt{locationd} downloads (Section \ref{sec:location_sent}).

Figure \ref{fig:wloc_cell} shows a cell tower identifier sent to \texttt{wloc}.

\begin{figure}[htb]
\centering
\begin{lstlisting}[frame=single,language=http]
POST https://gs-loc.apple.com/clls/wloc
!{\color{blue}Request Headers:}!
User-Agent : locationd/2787.0.5 CFNetwork/1410.1 Darwin/22.6.0
!{\color{blue}POST Body (decoded):}!
apple arpc header: version=1, locale=en_US, app=com.apple.locationd, osVersion=16.7.11.20H360, requestType=1, length=66
numberOfSurroundingWifis: 1
lteCellTowers {
  mcc: 272, mnc: 5
  cellID: 1401601, tacID: 40601
}
surroundingWifiBands: K2DOT4_GHZ
wifiAltitudeScale: K_WIFI_ALTITUDE_SCALE10TO_THE2
meta {softwareBuild: "iPhone OS16.7.11/20H360", productId: "iPhone10,1"}
!{\color{blue}Response Data:}!
lteCellTowers {
  mcc: 272, mnc: 5
  cellID: 1401601, tacID: 40601
  location {
    latitude: 53.31<redacted>, longitude: -6.29<redacted>
    accuracy: 1414, score: 937, reach: 1554
  }
}
<more cell tower identifiers, their lat/lon locations and coverage>
\end{lstlisting}
\caption{Example showing transmission of a cell tower identifier to Apple server \nolinkurl{gs-loc.apple.com/clls/wloc}.}\label{fig:wloc_cell}
\end{figure}

Figure \ref{fig:rtloc} shows the real-time traffic probe sent by \texttt{geod}. It carries the same GPS fixes as the \texttt{/pds/pd} upload in Figure \ref{fig:pdspd}.

\begin{figure}[htb]
\centering
\begin{lstlisting}[frame=single,language=http]
POST https://gsp9-ssl.apple.com/hvr/v2/rtloc
!{\color{blue}Request Headers:}!
User-Agent : geod/1 CFNetwork/1410.1 Darwin/22.6.0
!{\color{blue}POST Body (decoded):}!
apple arpc header: version=2, locale=en-IE_IE, app=com.apple.geod, osVersion=16.7.11.20H360, compressionAlgo=LZ4, requestType=1017, length=2376
logMessages {
  logMsgEvents {
    logMsgStates {
      stateType: DEVICE_IDENTIFIER
      deviceIdentifier {deviceOsVersion: "16.7.11.20H360", deviceHwIdentifier: "iPhone10,1"}
    }
    eventType: REALTIME_TRAFFIC_PROBE_COLLECTION
    realtimeTrafficProbeCollection {
      locationCollection {
        meta {productId: "D20AP", softwareVersion: "iPhone OS16.7.11/20H360"}
        locations {
          latitude: 53.37<redacted>
          longitude: -6.59<redacted>
          horizontalAccuracy: 5, altitude: 61.95, verticalAccuracy: 4
          timestamp: 802283024.0005983
          provider: 1, appBundleIdIndices: "0"
          rawMotionActivity {type: IN_VEHICLE_STATIC, confidence: HIGH}
          motionActivity {type: IN_VEHICLE_STATIC, confidence: HIGH}
          dominantMotionActivity {type: IN_VEHICLE_STATIC, confidence: HIGH}
        }
        <similar entries with handset location at different times>
        tripId: "7B0447BD-<redacted>"
        appBundleIds: "com.apple.Maps"
      }
    }
  }
  logMessageType: REALTIME_TRAFFIC_PROBE
}
\end{lstlisting}
\caption{Example showing transmission of device GPS locations to Apple server \nolinkurl{gsp9-ssl.apple.com/hvr/v2/rtloc}.  The locations are carried in a serialised CoreLocationProtobuf message, shown decoded.  Enumeration names are shortened.}\label{fig:rtloc}
\end{figure}

\subsection{\texorpdfstring{Examples Of \texttt{geod} Network Connections}{Examples Of geod Network Connections}}\label{examples-of-geod-network-connections}

\begin{figure}[htb]
\centering
\begin{lstlisting}[frame=single,language=http]
POST https://gsp-ssl.ls.apple.com/ab.arpc
!{\color{blue}Request Headers:}!
User-Agent : geod/1 CFNetwork/1410.1 Darwin/22.6.0
!{\color{blue}POST Body (decoded):}!
apple arpc header: version=1, locale=en-IE_IE, app=com.apple.geod, osVersion=16.7.11.20H360, requestType=1016, length=83
guid: "53101232-<redacted>"
clientMetadata {
  deviceCountryCode: "IE", deviceDisplayLanguage: "en-IE", displayRegion: "IE"
  additionalEnabledMarkets {}
  resultListAttributionSupport: MULTIPLE_ATTRIBUTIONS
  hourOfDay: 11, dayOfWeek: 2
  requiredVersion: VERSION_3
  localizationCapabilities {supportsLocalizedTrafficControlIcons: true}
  clientRevisions: AZUL_RELEASE
  <...>
}
!{\color{blue}Response Data:}!
apple header: version=1, responseType=1016, payloadLength=80
requestGuid: "53101232-<redacted>"
refreshIntervalSeconds: 14400
clientConfig {}
parsecClientMetadata {clientDatasetMetadata {bucketId: 7428}}
siriClientMetadata {clientDatasetMetadata {bucketId: 7428}}
rapClientMetadata {clientConfigs {}, clientDatasetMetadata {bucketId: 7428}}
mapsAbClientMetadata {clientDatasetMetadata {bucketId: 7428}}
branchExpirationTtlHours: 720
\end{lstlisting}
\caption{Example showing a connection to the Apple A/B experiment endpoint (\texttt{ab\_experiment\_url}), \nolinkurl{gsp-ssl.ls.apple.com/ab.arpc}.}\label{fig:ab}
\end{figure}

\begin{figure}[htb]
\centering
\begin{lstlisting}[frame=single,language=http]
POST https://gsp64-ssl.ls.apple.com/hvr/v3/use
!{\color{blue}Request Headers:}!
User-Agent : geod/1 CFNetwork/1410.1 Darwin/22.6.0
!{\color{blue}POST Body (decoded):}!
apple arpc header: version=2, locale=en-IE_IE, app=com.apple.geod, osVersion=16.7.11.20H360, compressionAlgo=LZ4, requestType=1017
logMessages {
  logMsgEvents {
    logMsgStates {stateType: APPLICATION_IDENTIFIER, applicationIdentifier {appIdentifier: "com.apple.geod"}}
    <...>
    logMsgStates {stateType: DEVICE_CONNECTION, deviceConnection {deviceNetworkConnectivity: WIFI, deviceCountryCode: "IE", cellularDataState: ALLOWED}}
    logMsgStates {stateType: DEVICE_IDENTIFIER, deviceIdentifier {deviceOsVersion: "16.7.11.20H360", deviceHwIdentifier: "iPhone10,1"}}
    logMsgStates {stateType: USER_SESSION, userSession {sessionId {low: 8071072902065579711, high: 21376328107327480}, ageOfSessionIdInSeconds: 432000}}
    eventType: NETWORK
    networkEvent {
      networkService: WIFI_QUALITY
      requestErrorDomain: "NSURLErrorDomain", requestErrorCode: -1004
      requestAppIdentifier: "com.apple.wifi.ThreeBarsXPCService"
      totalTime: 186, tilesetId: 2602162
    }
  }
  logMessageType: LONG_SESSION_USAGE
}
\end{lstlisting}
\caption{Example showing transmission of telemetry data to Apple server \nolinkurl{gsp64-ssl.ls.apple.com/hvr/v3/use}.}\label{fig:telemetry}
\end{figure}

\subsection{\texorpdfstring{HTTP Headers And Query Parameters In \texttt{geod} Connections}{HTTP Headers And Query Parameters In geod Connections}}\label{http-headers-and-query-parameters-in-geod-connections}

\begin{table*}[htb]
\footnotesize
\centering
\caption{HTTP query parameters observed in \texttt{geod} network traffic.}\label{tab:geod_query_params}
\vspace{4pt}
\begin{tabular}{|p{4.75cm}|l|l|p{6.25cm}|}
\hline
\textbf{Endpoint} & \textbf{Parameter} & \textbf{Example value} & \textbf{Description} \\
\hline
\multirow{3}{5.5cm}{\texttt{configuration.ls.apple.com}\newline\texttt{/config/defaults}}
  & \texttt{os}            & \texttt{ios}          & Operating system identifier \\
\cline{2-4}
  & \texttt{os\_version}   & \texttt{16.7.11}      & OS version string \\
\cline{2-4}
  & \texttt{hardware}      & \texttt{iPhone10,1}   & Hardware model identifier \\
\hline
\multirow{7}{5.5cm}{\texttt{gspe35-ssl.ls.apple.com}\newline\texttt{/geo\_manifest/dynamic/config}}
  & \texttt{application}          & \texttt{geod}       & Client application name \\
\cline{2-4}
  & \texttt{application\_version} & \texttt{1}          & Application version \\
\cline{2-4}
  & \texttt{country\_code}        & \texttt{IE}         & Country code \\
\cline{2-4}
  & \texttt{hardware}             & \texttt{iPhone10,1} & Handset model \\
\cline{2-4}
  & \texttt{os}                   & \texttt{ios}        & Operating system \\
\cline{2-4}
  & \texttt{os\_build}            & \texttt{20H360}     & OS build \\
\cline{2-4}
  & \texttt{os\_version}          & \texttt{16.7.11}    & OS version  \\
\hline
\multirow{10}{5.5cm}{\texttt{gspe19-ssl.ls.apple.com}\newline\texttt{/tile.vf}}
  & \texttt{style}      & \texttt{62, 68, 73}         & Map tile style ID \\
\cline{2-4}
  & \texttt{size}       & \texttt{2}                  & Tile size (Retina factor) \\
\cline{2-4}
  & \texttt{scale}      & \texttt{0}                  & Display scale \\
\cline{2-4}
  & \texttt{v}          & \texttt{18946531}           & Map data version number \\
\cline{2-4}
  & \texttt{z}          & \texttt{2, 9, 12}           & Zoom level (XYZ tile scheme) \\
\cline{2-4}
  & \texttt{x}          & \texttt{1, 246, 1976}       & Tile X coordinate \\
\cline{2-4}
  & \texttt{y}          & \texttt{1, 165, 1328}       & Tile Y coordinate \\
\cline{2-4}
  & \texttt{vlang}      & \texttt{en-GB}              & Display language\\
\cline{2-4}
  & \texttt{preflight}  & \texttt{2}                  & Request token \\
\cline{2-4}
  & \texttt{accessKey}  & \textit{Base64 token}       & $-$ \\
\hline
\multirow{3}{5.5cm}{\texttt{gspe19-ssl.ls.apple.com}\newline\texttt{/asset/v3/material/<id>}}
  & \texttt{style}      & \texttt{60}                & Map tile style ID \\
\cline{2-4}
  & \texttt{preflight}  & \texttt{2}                 & Request token \\
\cline{2-4}
  & \texttt{accessKey}  & \textit{Base64 token}      & $-$ \\
\hline
\multirow{1}{5.5cm}{\texttt{gspe21-ssl.ls.apple.com}\newline\texttt{/\{type\}/\{resource\}}}
  & \texttt{accessKey}  & \textit{Base64 token}       & Used for all static asset types: \texttt{font}, \texttt{icon}, \texttt{texture}, \texttt{stylesheet}, \texttt{xml}, \texttt{other} \\
\hline
\end{tabular}
\end{table*}

\begin{table*}[!htb]
\centering
\footnotesize
\caption{Non-standard HTTP headers observed in \texttt{geod} network connections. Standard headers (\texttt{Accept}, \texttt{User-Agent},
etc.) are omitted.}
\label{tab:geod_headers}
\begin{tabular}{|p{4.2cm}|p{9.5cm}|}
\hline
\textbf{Header} &
\textbf{Description / example value} \\
\hline
\texttt{maps-auth-token}           & HMAC of the tile version \texttt{v} (e.g.\ \texttt{v=18965870}); see Section \ref{sec:access-key} \\
\hline
\texttt{x-tilekey}                 & Integer tile key, e.g.\ \texttt{81607936}          \\
\hline
\texttt{maps-tile-x / -y / -z}     & XYZ tile grid coordinates (column, row, zoom level), e.g.\ \texttt{x=3, y=2, z=3}  \\
\hline
\texttt{maps-tile-style}           & Tile parameters, e.g.\ \nolinkurl{style=20&size=2&scale=0&v=18965870}.  The headers carry the same fields as the \texttt{tile.vf} query parameters of Table~\ref{tab:geod_query_params}; styles 1, 11, 20, 30 and 53 and zoom levels 3 to 15 occur in them  \\
\hline
\texttt{x-msg-priority}            & Small integer, either \texttt{5} or  \texttt{8}                        \\
\hline
\texttt{x-os-version}              & iOS version and build string, \texttt{16.7.11.20H360}                  \\
\hline
\texttt{maps-tile-reason}          & Fixed value, \texttt{PROACTIVE}                          \\
\hline
\end{tabular}
\end{table*}
\end{document}